\documentclass [12pt]{article}

\pdfoutput=1

\usepackage{a4wide, mathdots, amsmath, amsfonts, amssymb, geometry}
\usepackage{bm}
\usepackage{fancyhdr}
\usepackage{mathrsfs}  
\usepackage[latin1]{inputenc}
\usepackage[pdftex]{graphicx} 
\usepackage{graphicx}
\usepackage{float}
\usepackage{type1cm}
\usepackage{color}
\definecolor{dgreen}{rgb}{0,0.70,0.30}
\definecolor{gold}{rgb}{0.85,.66,0}
\definecolor{purple}{rgb}{1.0,0.3,0.6}
\usepackage{setspace} 
\usepackage{cite}

\usepackage{tikz}
\usetikzlibrary{calc} \usetikzlibrary{patterns} \usetikzlibrary{decorations.pathreplacing} \usetikzlibrary{decorations.markings} \usetikzlibrary{decorations.pathmorphing} \usetikzlibrary{positioning}

\restylefloat{figure}
\usepackage{epstopdf}

\def\Tijk#1,#2,#3{T^i_{#1}T^j_{#2}T^k_{#3}}

\newcommand{\nwc}{\newcommand}
\nwc{\ba}  {\begin{array}}
\nwc{\ea}  {\end{array}}
\nwc{\bdm} {\begin{displaymath}}
\nwc{\edm} {\end{displaymath}}

\nwc{\bea} {\begin{equation}\ba{lcl}}
\nwc{\eea} {\ea\end{equation}}

\nwc{\bda} {\bdm\ba{lcl}} 
\nwc{\eda} {\ea\edm}

\nwc{\bc}  {\begin{center}}
\nwc{\ec}  {\end{center}}

\nwc{\ds}  {\displaystyle}
\nwc{\bmat}{\left(\ba}
\nwc{\emat}{\ea\right)}
\nwc{\nn}  {\nonumber}
\nwc{\nnn} {\nonumber \vspace{.2cm} \\ }
\nwc{\ra}  {\rightarrow}
\nwc{\lra} {\longrightarrow}

\nwc{\p} {\partial}
\nwc{\num} {\mathfrak{n}}

\def\beq{\begin{equation}}
\def\eeq{\end{equation}}

\def\Re{{\rm Re\,}}
\def\Im{{\rm Im\,}}

\newcommand{\vecb}{\left(\begin{array}{c}}
\newcommand{\vece}{\end{array}\right)}
\newcommand{\ccb}{\left(\begin{array}{cc}}
\newcommand{\cce}{\end{array}\right)}
\newcommand{\cccb}{\left(\begin{array}{ccc}}
\newcommand{\ccce}{\end{array}\right)}
\newcommand{\ccccb}{\left(\begin{array}{cccc}}
\newcommand{\cccce}{\end{array}\right)}
\newcommand{\cccccb}{\left(\begin{array}{ccccc}}
\newcommand{\ccccce}{\end{array}\right)}

\newcommand{\pa}{\partial}

\newcommand{\de}{\delta}

\newcommand{\Ga}{\Gamma}

\newcommand{\te}{\textrm}

\newcommand{\co}{\ , \ \ \ \ \ \ }
\newcommand{\dd}{\mathrm{d}}

\newcommand{\ap}{\alpha'}

\newcommand{\MHV}{\Big|_{\textrm{MHV}}^{1^{-} 2^{-}}}

\begin{document}

\title{\textbf{String-inspired BCJ numerators} \\
\textbf{for one-loop MHV amplitudes}\\ \small{\phantom{.}} }
\author{\large Song He$^{\te{a,b,c}}$, Ricardo Monteiro$^{\te{d}}$, and 
Oliver Schlotterer$^{\te{e}}$\\[2cm]}
\date{}
\maketitle
\vskip-2cm

\centerline{\it $^{\te{a}}$ State Key Laboratory of Theoretical Physics and Kavli Institute for Theoretical Physics China,}
\centerline{\it Institute of Theoretical Physics, Chinese Academy of Sciences, Beijing 100190, P. R. China,}
\vskip0.15cm
\centerline{\it $^{\te{b}}$ Perimeter Institute for Theoretical Physics,}
\centerline{\it Waterloo, ON N2L 2Y5, Canada,}
\vskip0.15cm
\centerline{\it $^{\te{c}}$ School of Natural Sciences, Institute for Advanced Study,}
\centerline{\it Princeton, NJ 08540, USA, }
\vskip0.15cm
\centerline{\it $^{\te{d}}$ Mathematical Institute, University of Oxford,}
\centerline{\it 
Oxford OX2 6GG, UK,}
\vskip0.15cm
\centerline{\it $^{\te{e}}$ Max-Planck-Institut f\"ur Gravitationsphysik,} 
\centerline{\it Albert-Einstein-Institut, 14476 Potsdam, Germany.}

\bigskip
\begin{abstract}
We find simple expressions for the kinematic numerators of one-loop MHV amplitudes in maximally supersymmetric Yang--Mills theory and supergravity, at any multiplicity. The gauge-theory numerators satisfy the Bern-Carrasco-Johansson (BCJ) duality between color and kinematics, so that the gravity numerators are simply the square of the gauge-theory ones. The duality holds because the numerators can be written in terms of structure constants of a kinematic algebra, which is familiar from the BCJ organization of self-dual gauge theory and gravity. The close connection that we find between one-loop amplitudes in the self-dual case and in the maximally supersymmetric case is reminiscent of the dimension-shifting formula. The starting point for arriving at our expressions is the dimensional reduction of ten-dimensional amplitudes obtained in the field-theory limit of open superstrings.
 \end{abstract}

\maketitle{}

\setcounter{tocdepth}{2}

\newpage

\linespread{1.2}
\tableofcontents
\linespread{1.4}

\numberwithin{equation}{section}



\section{Introduction}   

Recent progress on scattering amplitudes has sought to describe these quantities in a variety of manners, which highlight different properties and symmetries  of a theory, or connections between different theories. The Bern-Carrasco-Johansson (BCJ) duality between color and kinematics \cite{Bern:2008qj} resulted from exploring the connections between amplitudes in gauge theory and in gravity. Refs.~\cite{Bern:2008qj,Bern:2010ue} found that a gauge-theory amplitude can often be expressed such that its kinematic dependence closely mirrors its color dependence. The kinematic numerators corresponding to trivalent diagrams are known as \emph{BCJ numerators} when they satisfy this ``duality" between color and kinematics. Their most remarkable property is that gravity amplitudes can be obtained in a simple manner from the gauge-theory ones by substituting color factors for another copy of BCJ numerators. This procedure to construct gravity amplitudes has led to great advances in the study of the ultraviolet behaviour of supergravity theories~\cite{Bern:2010ue,Bern:2012uf,Bern:2012cd,Bern:2013qca,Bern:2013uka,Bern:2014sna}.

The BCJ duality remains a conjecture at loop level, and the principles at work are poorly understood. Most loop-level results have followed from an \emph{ad hoc} approach to the construction of BCJ numerators, starting with an ansatz and fixing it by unitarity cuts. The difficulty in extending this approach to higher loops is a major obstacle at present. In this work, we bring together two lines of research which explore the mathematical structure underlying the BCJ duality. As a result, we find simple and suggestive expressions for MHV amplitudes at one loop, in maximally supersymmetric theories.

One line of research on the principles behind BCJ is based on string theory. Amplitudes of various field theories can be obtained in the infinite tension limit ($\alpha'\to 0$) of superstring amplitudes~\cite{Green:1982sw}. The meta-structure provided by string theory has already led to many insights and practical applications in the study of field-theory amplitudes, starting with the Kawai-Lewellen-Tye (KLT) relations between gauge and gravity amplitudes at tree level~\cite{Kawai:1985xq}, and the BCJ story is no exception. The color-kinematics duality in gauge theory leads to linear relations between tree-level color-ordered amplitudes, known as BCJ relations \cite{Bern:2008qj}. The first proof of these relations was based on the monodromy properties of integrals appearing in string amplitudes \cite{BjerrumBohr:2009rd,Stieberger:2009hq}. Moreover, the first explicit local expressions for BCJ numerators at tree level were derived in ref.~\cite{Mafra:2011kj} from the pure-spinor formulation of superstring theory \cite{Berkovits:2000fe}. This line of work has been streamlined and generalized in the framework of multiparticle superfields \cite{Mafra:2014oia, Mafra:2015gia} which has led to important insights on the BCJ conjecture at loop level \cite{Mafra:2014gja,Mafra:2015mja}. These building blocks also allow one to determine one-loop amplitudes of the ten-dimensional open superstring \cite{Mafra:2012kh}, or at least their BRST-invariant subsector which is unaffected by the gauge anomaly \cite{Green:1984sg, Green:1984qs}. Constraining the helicities of their four-dimensional reduction to MHV configurations is used as a driving force to derive the all-multiplicity BCJ numerators presented in this work.

Another line of research on BCJ is based on the direct search for a ``kinematic algebra" which mirrors the color Lie algebra, making the duality between color and kinematics manifest. Such a kinematic algebra has indeed been found for the self-dual sector of gauge theory and gravity \cite{Monteiro:2011pc}. At tree level, while amplitudes in the self-dual theories vanish, this structure is essentially preserved for the closely-related MHV amplitudes. At one loop, the amplitudes in the (non-supersymmetric) self-dual theories do not vanish, and the kinematic algebra allows for the construction of BCJ numerators for all-plus amplitudes, and also for the closely related one-minus amplitudes \cite{Boels:2013bi}. These are the only two families of loop-level amplitudes for which explicit BCJ numerators were known for any multiplicity. Our results add one more all-multiplicity family, that of MHV amplitudes in maximally supersymmetric gauge theory and gravity. While five-point BCJ numerators~\cite{Carrasco:2011mn} and a procedure to implicitly determine them for higher multiplicities~\cite{Bjerrum-Bohr:2013iza} were known, we present here closed-form and remarkably simple expressions for any multiplicity. A notable feature of our BCJ numerators is that the only poles appearing in them can be traced back to reference spinors of polarization vectors; our BCJ numerators (\ref{massgen}) are manifestly local after factoring out simple prefactors. 

We show that there is a direct connection between one-loop MHV amplitudes obtained from string theory, and the kinematic algebra of the self-dual theories. Our numerators (\ref{massgen}), provided by the field-theory limit of strings, can be conveniently presented in terms of kinematic structure constants. We identify a prescription $\mathscr{X}$, which maps the kinematic structure constants in the self-dual numerators to our BCJ numerators for MHV amplitudes in maximally supersymmetric theories. This connection is reminiscent of the result of ref.~\cite{Bern:1996ja}, which gives one-loop all-plus amplitudes in non-supersymmetric gauge theory and gravity (which correspond to the self-dual sector) in terms of a dimension-shifting rule applied to the integrand of one-loop MHV amplitudes in the maximally supersymmetric theories. The simple deformation rule we found suggests that there is a wider story to explore, and one might speculate about generalizations to arbitrary helicities, higher dimensions and even higher loop orders. 

This paper is organized as follows: in section~\ref{sec2}, we briefly review the BCJ duality and kinematic structure constants in the self-dual sector. Section~\ref{sec3} contains the construction of tree-level BCJ numerators, with the example of MHV amplitudes, both from the self-dual kinematic algebra and from the field-theory limit of string theory. We present the general results for BCJ numerators for one-loop MHV amplitudes in section~\ref{sec4}, highlighting the connection with self-dual numerators and the relation to the dimensional-shifting formula. In section~\ref{sec5}, we show how these results are derived from string theory at one loop. We end with conclusions and outlook in section~\ref{sec6}.

\section{Review}\label{sec2}

\subsection{BCJ duality and double copy}

\begin{figure}[h]
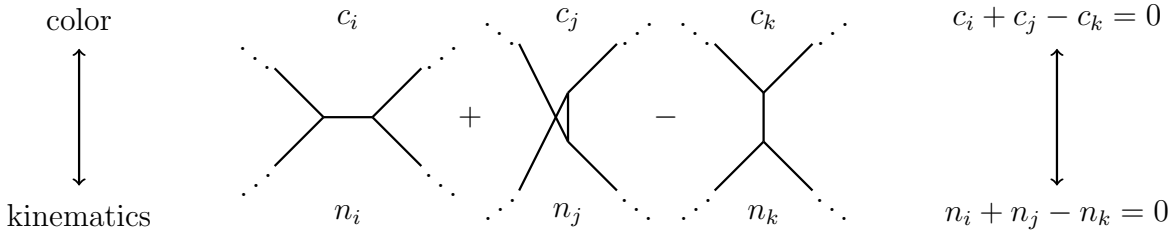

\begin{center}
 \tikzpicture [scale=1.3]
 \draw (-3,-1) node {kinematics};
 \draw (-3,1) node {color};
 \draw [line width=0.30mm,<->] (-3,0.7) -- (-3,-0.7);
 \draw (7,-1) node { $n_i + n_j - n_k= 0$};
 \draw (7,1) node { $c_i + c_j - c_k= 0$};
 \draw [line width=0.30mm,<->] (7,0.7) -- (7,-0.7);
 \scope[yshift=-0.5cm, xshift=-2.5cm]
 \draw [line width=0.30mm]  (2,0.5) -- (1.5,1) ;
 \draw (1.3,1.2) node {$\ddots$};
 \draw [line width=0.30mm]  (2,0.5) -- (1.5,0) ;
 \draw (1.3,-0.1) node {$\iddots$};
 \draw [line width=0.30mm]  (2,0.5) -- (2.5,0.5) ;
 \draw [line width=0.30mm]  (2.5,0.5) -- (3,1) ;
 \draw (3.2,1.2) node {$\iddots$};
 \draw [line width=0.30mm]  (2.5,0.5) -- (3,0) ;
 \draw (3.2,-0.1) node {$\ddots$};
 \draw (2.25,-0.5) node {$\displaystyle  n_i$}; 
 \draw (2.25,1.5) node {$\displaystyle c_i$}; 
 \endscope
 \scope[xshift=-3.5cm]
 \draw (4.5,0) node{$+$};
 \draw [line width=0.30mm]  (5.5,-0.25) -- (5,0.75) ;
 \draw (4.8,0.95) node {$\ddots$};
 \draw [line width=0.30mm]  (5.5,0.25) -- (5,-0.75) ;
 \draw (4.8,-0.85) node {$\iddots$};
 \draw [line width=0.30mm]  (5.5,0.25) -- (5.5,-0.25) ;
 \draw [line width=0.30mm]  (5.5,0.25) -- (6,0.75) ;
 \draw (6.2,0.95) node {$\iddots$};
 \draw [line width=0.30mm]  (5.5,-0.25) -- (6,-0.75);
 \draw (6.2,-0.85) node {$\ddots$};
 \draw (5.5,-1) node {$\displaystyle n_j$}; 
\draw (5.5,1) node {$\displaystyle c_j$}; 
 \endscope
 \scope[xshift=-1.5cm]
 \draw (4.5,0) node{$-$};
 \draw [line width=0.30mm]  (5.5,0.25) -- (5,0.75) ;
 \draw (4.8,0.95) node {$\ddots$};
 \draw [line width=0.30mm]  (5.5,-0.25) -- (5,-0.75) ;
 \draw (4.8,-0.85) node {$\iddots$};
 \draw [line width=0.30mm]  (5.5,0.25) -- (5.5,-0.25) ;
 \draw [line width=0.30mm]  (5.5,0.25) -- (6,0.75) ;
 \draw (6.2,0.95) node {$\iddots$};
 \draw [line width=0.30mm]  (5.5,-0.25) -- (6,-0.75);
 \draw (6.2,-0.85) node {$\ddots$};
 \draw (5.5,-1) node {$\displaystyle n_k$}; 
 \draw (5.5,1) node {$\displaystyle  c_k$}; 
 \endscope
 \endtikzpicture
\end{center}
\caption{A diagrammatic representation of color and kinematic Jacobi relations.}
\label{fig4}
\end{figure}

The BCJ duality, or color-kinematics duality, states that the kinematic dependence of a gauge-theory amplitude can be expressed so that it has the same algebraic properties as the color dependence \cite{Bern:2008qj,Bern:2010ue}. In order to present it precisely, we first write down the $N$-point amplitude at a given loop order $L$ in terms of a sum over trivalent diagrams, i.e.~diagrams with cubic vertices,
\beq
\label{LoopGauge}
A_N^{L-\te{loop}} = \sum_{\te{diagrams }\Gamma_i}\int \prod_{j=1}^L \frac{\dd^D\ell_j}{(2\pi)^D}\frac{1}{S_i}\frac{n_i(\ell)\,c_i}{\prod_k P_{k,i}(\ell)} \ .
\eeq
For each trivalent diagram $\Ga_i$, $S_i$ is the symmetry factor (to avoid overcounting), and $P_{k,i}^{-1}$ are the propagators, which generically depend on loop momenta $\ell_j$. The color factors $c_i$ can be straightforwardly read off from the graph $\Gamma_i$ by dressing each cubic vertex with a structure constant $f^{abc}$ of the gauge group, and $n_i$ are kinematic numerators depending on polarizations and (internal or external) momenta. It is clearly always possible to write down the amplitude in this manner, as four-point vertices can be decomposed according to the color structures and absorbed into the trivalent diagrams by changing $n_i$. In fact, there is a large ambiguity in the choice of kinematic numerators, since the color factors are not independent. These originate from the Lie algebra of the gauge group, and thus satisfy Jacobi identities of the type
$$c_i + c_j - c_k=0 \ .$$
For instance, at tree level with four particles ($N=4$), there are three trivalent diagrams, and their color factors satisfy the identity:\, $f^{a_1a_2b} f^{ba_3a_4} + f^{a_1a_3b} \, f^{ba_4a_2} - f^{a_1a_4b} \, f^{ba_3a_2}  =  0$, with indices $a_i,b$ in the adjoint representation of the gauge group. The precise statement of the BCJ duality is that there exists a choice of kinematic numerators that satisfy the same algebraic identities as the associated color factors for each value of the loop momenta $\ell$,
\beq
n_i(\ell) + n_j(\ell) - n_k(\ell) =0  \ . \label{jacob}
\eeq
See figure~\ref{fig4} for a diagrammatic representation of these identities.

The BCJ double copy states that, if such a representation of the gauge-theory amplitude is available, then a gravity amplitude is obtained straightforwardly, substituting the color factors by another copy of the kinematic numerators \cite{Bern:2010yg},
\beq
\label{LoopSq}
M_N^{L-\te{loop}} = \sum_{\te{diagrams }\Gamma_i}\int \prod_{j=1}^L \frac{\dd^D\ell_j}{(2\pi)^D}\frac{1}{S_i}
\frac{n_i(\ell)\tilde n_i(\ell)}{\prod_k P_{k,i}(\ell)}=A_N^{L-\te{loop}}|_{c_i \to \tilde n_i (\ell)}\ .
\eeq
The states involved in the gravity scattering are the direct product of the states involved in the gauge-theory scattering. For instance, if the numerators $n_i$ ($\tilde n_i$) correspond to gluonic states with polarization vectors $\epsilon^\mu_i$ ($\tilde \epsilon^\mu_i$), then the gravity states correspond to the polarization tensors $\varepsilon_i^{\mu\nu} = \epsilon^\mu_i \tilde \epsilon^\nu_i$, which can in general be decomposed into graviton, dilaton and B-field components.

Both the color-kinematics duality and the double copy are well understood at tree level~\cite{Bern:2010yg,BjerrumBohr:2010zs,Feng:2010my,Tye:2010dd,Mafra:2011kj,Monteiro:2011pc,BjerrumBohr:2012mg,Fu:2012uy}. 
The recently developed formalism of the scattering equations has brought a new insight into these structures~\cite{Cachazo:2013gna,Cachazo:2013hca,Cachazo:2013iea,Litsey:2013jfa,Monteiro:2013rya,Mason:2013sva}.  At loop level, examples of amplitudes which admit a BCJ form have been presented in refs.~\cite{Bern:2010ue,Carrasco:2011mn,Carrasco:2012ca,Bern:2012uf,Boels:2013bi,Bjerrum-Bohr:2013iza,Bern:2013yya,Bern:2013qca,Nohle:2013bfa,Bern:2013uka,Ochirov:2013xba,Chiodaroli:2013upa,Mafra:2014oia,Bern:2014sna,Chiodaroli:2014xia,Mafra:2014gja,Mafra:2015mja, Ellis}.
While there is all-loop evidence in certain kinematic limits~\cite{Oxburgh:2012zr,Saotome:2012vy}, the existence of BCJ numerators for any multiplicity and loop order remains conjectural.
Generalizations of the BCJ structure have been studied in various contexts, including the introduction of massive quarks \cite{Johansson:2015oia} and the application to Chern-Simons theories \cite{Bargheer:2012gv,Huang:2012wr}. An extension of the perturbative relations between gauge theory and gravity to classical solutions was initiated in refs.~\cite{Monteiro:2014cda, Luna:2015paa}; see also ref.~\cite{Anastasiou:2014qba}.
Ref.~\cite{Carrasco:2015iwa} provides a recent comprehensive review.

\subsection{Self-dual gauge theory and gravity}

In four dimensions, both gauge theory and gravity can be truncated to their self-dual sectors, resulting in much simpler theories. It was shown in ref.~\cite{Monteiro:2011pc} that the BCJ structure can be made completely manifest in the self-dual theories.

Let us focus on the gauge-theory case. We can write down the Feynman rules in light-cone gauge, with the light-cone defined by the null vector $\eta$. In spinor-helicity language \cite{Boels:2013bi},

$\bullet$ vertices: $\pm$ denote the helicities, and $i_\eta \equiv 2\,k_i\cdot \eta= \langle\eta|k_i|\eta]=  \langle\eta i \rangle [i \eta]$,
\begin{equation}\label{LCvertices}
\begin{gathered}
(i^+,j^+,l^-)=\frac{l_\eta}{i_\eta j_\eta}\, X_{i,j} \, f^{a_i a_j a_l}, \qquad \textrm{with} \quad X_{i,j} \equiv \langle\eta| k_i k_j |\eta\rangle \ ,\\
(i^-,j^-,l^+)=\frac{l_\eta}{i_\eta j_\eta}\, \overline{X}_{i,j} \, f^{a_i a_j a_l}, \qquad \textrm{with} \quad \overline{X}_{i,j} \equiv [\eta| k_i k_j|\eta] \ ,\\
(i^+,j^+,l^-,m^-)=i \left( \frac{i_\eta l_\eta + j_\eta m_\eta}{(i_\eta+m_\eta)^2} \, f^{a_i a_m b} f^{b a_j a_l}
+ \frac{i_\eta m_\eta + j_\eta l_\eta}{(i_\eta+l_\eta)^2} \, f^{a_i a_l b} f^{b a_j a_m} \right) \ ,
\end{gathered}
\end{equation}

$\bullet$ propagators: \,$i\,\delta^{a_i a_j}/k^2$,

$\bullet$ external state factors: \quad $\displaystyle{\hat e_i^{(+)}=\frac{[\eta i]}{\langle \eta i\rangle} \ , \qquad \hat e_i^{(-)}=\frac{\langle \eta i\rangle} {[\eta i]}} \ $. \\

The quantities $X$ and $\overline X$ defined above using the spinor-helicity formalism are antisymmetric in their indices. In fact, they are simple spinor brackets, e.g.~$X_{i,j}=-[\![i,j]\!]$, where the spinor $|i ] \! ] \equiv k_i |\eta \rangle$ can be associated to off-shell momenta $k_i$. Gauge invariance guarantees that all physical quantities are independent of the choice of null vector $\eta$. The Feynman rules above follow from a light-cone action \cite{Chalmers:1998jb}; see ref.~\cite{Broedel:2011ib} for a supersymmetric extension.

Self-dual gauge theory can be defined as the restriction of the interactions to the vertex ($+,+,-$) associated to $X$. A simple counting argument shows that the only helicity configurations allowed in the self-dual sector are one-minus ($-,+,+,\cdots,+$) at tree level, and all-plus ($+,+,+,\cdots,+$) at one loop. As it is well-known, the one-minus amplitudes vanish at tree level (see next subsection). This definition of self-dual gauge theory is equivalent to the common definition in terms of an equation of motion imposing the self-duality of the field strength, but has the advantage of extending it beyond tree level. The cubic vertex ($+,+,-$) can be seen as arising from the equation of motion in the light-cone gauge \cite{Bardeen:1995gk,Chalmers:1996rq}.

Analogous statements are valid in the gravity case. The equation of motion in pure self-dual gravity is the self-duality of the Riemann tensor, and a gauge choice reduces the classical problem to a scalar with a cubic interaction \cite{Plebanski:1975wn}. The discussion of helicities above also holds in the gravity case. Notice, however, that the double copy of self-dual gauge theory is the self-dual theory of graviton--dilaton--B-field.

\subsection{Kinematic structure constants}

We will now review how the object $X$ in eq.~(\ref{LCvertices}) gives rise to a kinematic algebra which explains the BCJ duality in self-dual gauge theory and gravity \cite{Monteiro:2011pc}.
 
In the self-dual sector, the Feynman rules can be further simplified. In particular, we can strip off the prefactor $l_\eta(i_\eta j_\eta)^{-1}$ from the vertex ($+,+,-$) in eq.~\eqref{LCvertices}. These prefactors cancel along any diagram because each internal line has opposite helicities at its ends. They require only that the external factors be modified, $\hat e_{i}^{(\pm)}\rightarrow e_{i}^{(\pm)}$. The rules for self-dual gauge theory are then
\beq
(i^+,j^+,l^-)^{\text{s.d.}}=X_{i,j} \, f^{a_i a_j a_l}, \qquad e_i^{(+)}=-\frac{1}{\langle \eta i\rangle^2}, \qquad e_i^{(-)}=-\langle \eta i\rangle^2 \ ,
\label{extstates}
\eeq
whereas for self-dual gravity they are\footnote{We write the rules for gravitons only, but the dilaton and B-field can also be external states. In four dimensions, they correspond to the combinations $(+-)\pm(-+)$. A property of graviton scattering at one loop is that, while these extra states run in the loop, the effect is merely a factor of two with respect to pure self-dual gravity \cite{Boels:2013bi}.}
\beq
(i^{++},j^{++},l^{--})^{\text{s.d.}}=X_{i,j}^2, \qquad e_i^{(++)}=\frac{1}{\langle \eta i\rangle^4}, \qquad e_i^{(--)}=\langle \eta i\rangle^4 \ .
\eeq
The BCJ double copy is direct in the rules above. The reason for this is that the BCJ duality between color and kinematics is manifest in self-dual gauge theory. First, notice that $X_{i,j} = - X_{j,i} = X_{i,i+j}$, so that $X$ is completely antisymmetric in the external legs of the vertex, $\{k_i,k_j,-k_i-k_j\}$, just as $f^{a_i a_j a_l}$. Moreover, $X$ satisfies the Jacobi-type identity
\begin{equation}
X_{i,j} X_{i+j,l} + X_{j,l} X_{j+l,i} + X_{l,i} X_{l+i,j} =0 \ .
\label{Schout}
\end{equation}
In fact, this follows from the Schouten identity, if we think of $X$ as a spinor bracket as in eq.~(\ref{LCvertices}). Indeed, $X$ is the structure constant of an area-preserving diffeomorphism algebra (see ref.~\cite{Monteiro:2011pc} for more details). To conclude, there is a kinematic algebra associated to $X$ which mirrors the color Lie algebra associated to $f^{a_i a_j a_l}$.

\begin{figure}[t]
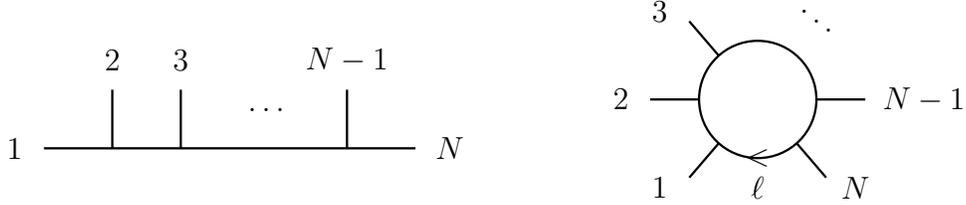

\begin{center}
 \tikzpicture [scale=1.3]
 \draw [line width=0.3mm] (-3.3,-0.5) -- (0.5,-0.5);
 \draw [line width=0.3mm] (-2.6,-0.5) -- (-2.6,0.1);
 \draw [line width=0.3mm] (-1.9,-0.5) -- (-1.9,0.1);
 \draw [line width=0.3mm] (-0.2,-0.5) -- (-0.2,0.1);
 \draw (-3.6,-.5) node {1}; 
 \draw (0.85,-.5) node {$N$}; 
 \draw (-2.6,0.4) node {2}; 
 \draw (-1.9,0.4) node {3}; 
 \draw (-0.2,0.4) node {$N-1$}; 
 \draw (-1,-0.1) node {$\dots$};
 \draw [line width=0.3mm] (4,0) circle (.6);
 \draw [line width=0.3mm] (2.9,0) -- (3.4,0);
 \draw [line width=0.3mm] (3.3,-.8) -- (3.6,-0.45);
 \draw [line width=0.3mm] (4.7,-.8) -- (4.4,-0.45);
 \draw [line width=0.3mm] (3.3,.8) -- (3.6,0.45);
 \draw [line width=0.3mm] (4.6,0) -- (5.1,0);
 \draw (2.6,0) node {2}; 
 \draw (3,-.9) node {1}; 
 \draw (3,.9) node {3}; 
 \draw (5,-.9) node {$N$}; 
 \draw (5.7,0) node {$N-1$}; 
 \draw (4.6,.9) node {$\ddots$};
 \draw(4,-0.6)node{$<$};
 \draw(4,-0.9)node{$\ell$}; 
 \endtikzpicture
\end{center}
\caption{The master topology is the half-ladder diagram at tree level (left) and the $N$-gon at one loop (right).}
\label{figmaster}
\end{figure}

Let us be concrete. At tree level, the self-dual sector corresponds to one-minus amplitudes. The BCJ numerator for the half-ladder topology  (see figure \ref{figmaster}) is
\begin{equation}
n^{\text{tree, s.d.}}_{1^+,2^+,\cdots,r^-,\cdots,N^+} = (-1)^N \langle \eta r\rangle^4 \left( \prod_{i=1}^N \frac{1}{\langle \eta i \rangle^2} \right) \prod_{j=2}^{N-1} X_{1+2+\cdots+(j-1),j}\ ,
\label{treecomb}
\end{equation}
where ``s.d." denotes the self-dual sector. The half-ladder is the master topology in the sense that, if its numerator is known, all other numerators can be obtained through Jacobi-type identities. It is trivial to see that these amplitudes vanish, since we can choose $|\eta\rangle=|r\rangle$ in the numerator (\ref{treecomb}).

At one loop, the self-dual sector corresponds to all-plus amplitudes. The master topology for $N$ particles is the $N$-gon (see figure \ref{figmaster}), and its BCJ numerator reads
\begin{equation}
\label{allplus}
n^{1-\text{loop, s.d.}}_{1^+|2^+|\cdots|N^+} = 2\, (-1)^N \prod_{i=1}^N \frac{1}{\langle \eta i \rangle^2} X_{\ell + 1+2+\cdots+(i-1),i}\ ,
\end{equation}
where we separate corners of the $N$-gon by vertical lines, and $\ell$ is the loop momentum; the factor of 2 follows from the two global choices of helicities for the internal lines.

We shall see that this structure is not exclusive to the self-dual sector. It was previously shown to underlie the BCJ duality for tree-level MHV (two-minus) amplitudes \cite{Monteiro:2011pc}, as we shall see in the next section, and for one-loop one-minus amplitudes; these are the closest helicity sectors to the ones discussed above (one more `minus' helicity). A generalization of the object $X$ to any tree-level amplitude, in any number of dimensions, has also been found \cite{Monteiro:2013rya}, based on the scattering equations. In the string-theory approach, the fingerprints of kinematic structure constants can be found in the OPEs among vertex operators. In the pure-spinor formalism \cite{Berkovits:2000fe}, this leads to the recursive formulae for multiparticle superfields \cite{Mafra:2014oia}.

\section{Tree-level numerators}\label{sec3}

This section, concerned with tree-level MHV amplitudes, prepares the ground for the one-loop results, and we check that two distinct procedures to get BCJ numerators give the same answer. The first, pursued in ref.~\cite{Mafra:2011kj}, is based on the field-theory limit of superstring amplitudes, obtained from the pure-spinor formalism. The second, pursued in ref.~\cite{Monteiro:2011pc}, is the use of light-cone Feynman rules, taking advantage of the close relationship between the MHV sector and the self-dual sector. 

We will present the derivation of tree-level MHV numerators from string theory in Appendix~\ref{app:string_tree}. Here we summarize some conventions:  the Mandelstam invariants are given by 
\beq
s_{ij} \equiv 2 (k_i \cdot k_j) = (k_i+k_j)^2 \co s_{i_1 i_2\ldots i_p } \equiv (k_{i_1} + k_{i_2}+\ldots +k_{i_p})^2 \ .
\label{mand}
\eeq
Upon dimensional reduction to $D=4$, we choose the following form for the polarizations,
\beq
\epsilon_i^{(+)}= \frac{|1\rangle [i|}{\langle i 1\rangle} \ , \qquad 
\epsilon_i^{(-)}= \frac{|i\rangle [\eta|}{[ i \eta ]} \ ,
\label{treenum5}
\eeq
i.e.~the reference spinors for positive and negative helicity particles are $|1 \rangle $ and $[\eta|$, respectively.

\subsection{Self-dual type numerators}

The MHV tree-level amplitudes are the same in pure Yang-Mills theory and in the gluon components of the maximally supersymmetric theory. Since at one loop we will be interested in the latter theory, we will use the supersymmetry formalism and leave the helicity choice of two negative-helicity gluons generic by working with the superamplitude; the supermomentum-conservation delta function $\delta^8(Q)$ becomes $\langle rs \rangle^4$, if $r$ and $s$ denote negative-helicity gluons. The BCJ numerators of half-ladder diagrams (see figure \ref{figmaster}) have been obtained in ref.~\cite{Monteiro:2011pc}, and we present them here in the spinor-helicity formalism:
\begin{equation}
n^{\text{tree,\,MHV}}_{\underline{1},2,\cdots,N} = \delta^8(Q) \, (-1)^N \left( \prod_{i=2}^N \frac{1}{\langle 1 i \rangle^2} \right)
\frac{\overline{X}_{1,2}}{[ \eta 1 ]^2}  \prod_{j=3}^{N-1} X_{1+2+\cdots+(j-1),j} \ , \qquad
\text{with} \quad |\eta\rangle = |1\rangle \ .
\label{treenumA}
\end{equation}
The special role of particle 1 could have been played by any other particle. This result is determined from the rules \eqref{LCvertices} if particle 1 is taken to have negative helicity (which is not required in the supersymmetric expression above). First, it can easily be shown by a counting argument that MHV amplitudes correspond to diagrams with one and only one non-($++-$) vertex, which may be a ($--+$) vertex or a four-point vertex. Second, by taking the gauge choice $|\eta\rangle \to |1\rangle$, all diagrams with a four-point vertex vanish, and in the remaining diagrams the ($--+$) vertex must be attached to particle 1, since $X_{1,i}\to 0$. We see now that by aligning the light-cone with one of the particles, we import for the MHV amplitude much of the structure of the self-dual theory. In order to make the BCJ properties completely transparent, we may also write the numerator only in terms of $X$ vertices:
\begin{equation}
n^{\text{tree,\,MHV}}_{\underline{1},2,\cdots,N} = \delta^8(Q) \,(-1)^N \left(\prod_{i=2}^N \frac{1}{\langle 1 i \rangle^2} \right) X_{q,2} \prod_{j=3}^{N-1} X_{1+2+\cdots+(j-1),j} \ , \quad
\ \ |\eta\rangle = |1\rangle \ , \ \  \;\; |\eta] = \frac{q|1\rangle}{[1\eta]} \ ,
\label{treenumB}
\end{equation}
where the momentum $q$ is defined by the last equation.

The relation between the famous Parke-Taylor formula \cite{Parke:1986gb} for MHV amplitudes, and the cubic-diagram expansion with numerators \eqref{treenumB}, can be seen from the identities
\beq
\frac{1}{\langle 12 \rangle \langle 23 \rangle \langle 31 \rangle} = \frac{-1}{\langle 12 \rangle^2 \langle 13 \rangle^2} \frac{ X_{2,3}}{s_{23}}
\co
\frac{1}{\langle 12 \rangle \langle 23 \rangle \langle 34 \rangle\langle 41 \rangle} = \frac{-1}{\langle 12 \rangle^2 \langle 13 \rangle^2\langle 14 \rangle^2} \left( 
 \frac{ X_{2,3} X_{2+3,4} }{s_{23} s_{234}} +   \frac{ X_{4,3} X_{4+3,2}  }{s_{34} s_{234}}
\right)
\label{massb3}
\eeq
and its all-multiplicity generalization
\beq
\frac{1}{\langle 12 \rangle \langle 23 \rangle \ldots \langle p{-}1,p \rangle \langle p1 \rangle} =  \frac{-1}{\langle 12 \rangle^2 \langle 13 \rangle^2 \cdots \langle 1p \rangle^2} \sum_{\te{cubic graphs with} \atop{\te{ordering}}\ \{2,3,\ldots,p\}} \frac{ \prod_{\te{vertices $j$ with} \atop{\te{edges $\{ a_j,b_j,-a_j-b_j\}$}}} X_{a_j,b_j} }{\prod_{\te{edges $k$}} s_k} \ .
\label{massb4}
\eeq
The sum over cubic graphs on the right hand side is known as a Berends-Giele current \cite{Berends:1987me} and designed to build up a color-ordered tree-amplitude, where an additional off-shell leg completes $\{2,3,\ldots,p\}$ to a cycle.

\section{One-loop numerators}\label{sec4}

We saw that tree-level MHV numerators are closely related to those in the self-dual theory, and that the representation which makes this clear can be obtained from string theory. In this section, we will state our main result -- 
all-multiplicity numerators for one-loop MHV amplitudes in ${\cal N}=4$ SYM which obey the BCJ duality. Their form in terms of kinematic structure constants shows that the one-loop MHV numerators in the maximally supersymmetric theory are also closely related to those in the self-dual theory (all-plus at one loop). After presenting the results, we will compare our findings to the so-called dimension-shifting formula \cite{Bern:1996ja}, which also relates these two families of one-loop amplitudes. These results were derived from the field-theory limit of superstring amplitudes, restricting to the four-dimensional MHV case.  We will leave the discussion on the string-theory derivation of the BCJ numerators to the next section.

\subsection{Notation and conventions}

\begin{figure}[t]
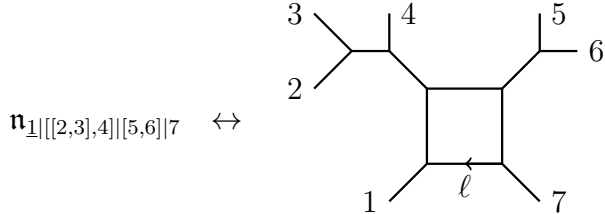

\begin{center}
\tikzpicture [scale=1,line width=0.30mm]
\draw (0,0) -- (1,0) ;
\draw (0,0) -- (0,1) ;
\draw (0,1) -- (1,1) ;
\draw (1,0) -- (1,1) ;
\draw[->](1,0)--(0.5,0) node[below]{$\ell$};
\draw (1,1) -- (1.5,1.5) ;
\draw (1.5,1.5) -- (1.5,2) node[right]{$5$} ;
\draw (1.5,1.5) -- (2,1.5) node[right]{$6$} ;
\draw (0,0) -- (-0.5,-0.5) node[left]{$1$};
\draw (0,1) -- (-0.5,1.5) ;
\draw(-0.5,1.5)--(-1,1.5);
\draw(-1,1.5) -- (-1.5,1)node[left]{$2$};
\draw(-1,1.5)-- (-1.5,2)node[left]{$3$};
\draw(-0.5,1.5)--(-0.5,2)node[right]{$4$};
\draw (1,0) -- (1.5,-0.5) node[right]{$7$} ;
\draw(-4,0.5)node{$\num_{\underline{1}| [[2,3],4] | [5,6] | 7} \ \ \leftrightarrow$};
\endtikzpicture
\end{center}
\caption{An example for our notation for one-loop numerators.}
\label{fig:expl}
\end{figure}

To simplify notation, we peel off universal prefactors from the numerators and focus on their non-universal parts $ \num_{\underline{1}|A_2|A_3|\ldots | A_m} (\ell)$ henceforth which we define by
\begin{equation}
n^{1-\text{loop,\;MHV}}_{\underline{1}|A_2|A_3|\ldots | A_m}(\ell) \equiv \frac{ \delta^8(Q) }{\langle 12 \rangle^2 \langle 13 \rangle^2\ldots  \langle 1N \rangle^2}\;\; \num_{\underline{1}|A_2|A_3|\ldots | A_m} (\ell) \ .
\label{defnot}
\end{equation}
The vertical bars in their subscripts separate different corners of a cubic $m$-gon diagram. Multiparticle labels $A_j$ refer to tree-level subdiagrams whose vertex structure is represented by bracketings as exemplified in figure \ref{fig:expl}. The relative factor between numerators $n^{1-\text{loop,\;MHV}}_{\ldots}$ and $\num_{\ldots}$ on the right hand side of eq.~(\ref{defnot}) is universal to any $N$-point diagram and therefore does not alter their BCJ properties. Hence, it is sufficient to show that the set of $\num_{\ldots}$ obeys the kinematic analogues (\ref{jacob}) of Jacobi relations.

Moreover, we will specialize $| \eta \rangle \rightarrow | 1 \rangle$ throughout the rest of this work, such that
\begin{equation}
X_{i,j} \equiv \langle 1| k_i  k_j | 1 \rangle \ ,
\label{defXX}
\end{equation}
with momenta $k_i$ and $k_j$ possibly off shell. As a consequence, diagrams with leg 1 attached to a massive corner (i.e.~a tree-level subdiagram involving non-trivial bracketings such as $[1,2]$) have a vanishing numerator; that is why the notation in eq.~(\ref{defnot}) dedicates an underlined standalone slot $\underline{1}|\ldots$ to this leg. Also, the choice $| \eta \rangle \rightarrow | 1 \rangle$ allows us to identify the ubiquitous denominator in eq.~(\ref{defnot}) involving all the $\langle 1j \rangle^2$ at $j\neq 1$ with external state factors as in eq.~(\ref{extstates}).

Furthermore, the loop momentum $\ell$ is always defined to reside in the edge preceding leg 1 in clockwise direction, see e.g.~figure \ref{fig:expl}. We will distinguish integrated single-trace subamplitudes\footnote{At one loop, multitrace subamplitudes in SYM are determined from their single-trace counterparts \cite{Bern:1994zx}.} from their integrands $I^{1-\te{loop}}_{1,2,\ldots,N}(\ell)$, and strip off the universal prefactor for MHV as in eq.~\eqref{defnot},
\begin{align}
A^{1-\te{loop}}(1,2,\ldots,N) &\equiv \int \frac{\dd^D\ell}{(2\pi)^D} \ I^{1-\te{loop}}_{1,2,\ldots,N}(\ell) \ , \notag \\ 
 I^{1-\te{loop, MHV}}_{1,2,\ldots,N}(\ell) &\equiv \frac{\delta^8(Q)}{\prod_{i=2}^N \langle 1 i\rangle^2} \ \mathcal{I}_{1,2,\ldots,N} (\ell) \ .
\label{subamp}
\end{align}
As we will see, $\mathcal{I}_{1,2,\ldots,N} (\ell) $ defined by eq.~(\ref{subamp}) can be entirely written in terms of propagators and kinematic structure constants (\ref{defXX}). Since any appearance of $\ell$ in the numerators occurs via $X_{\ell,\ldots}$, a shift of $\ell$ by the external momentum $k_1$ drops out and ensures the correct reflection properties of $m$-gons under $1,2,3,\ldots,m \rightarrow 1,m,\ldots,3,2$. Both the vanishing numerators with leg 1 in a massive $m$-gon corner and the convention for the loop momentum arise naturally in the string-theory setup described in section~\ref{sec:string}.

\subsection[Examples with $N=4,5,6$]{Examples with $\bm{N=4,5,6}$}

To warm up, we present here some examples of BCJ numerators for a small number of particles. The particular arrangement of $X$'s will be better understood when we present the general rule.

For $N=4$, the amplitude is simply a sum over boxes, with an overall coefficient given by \cite{Green:1982sw}
\begin{equation}
n^{1-\text{loop,\;MHV}}_{\underline{1}|2|3|4} = -\delta^8(Q) \frac{[12][34]}{\langle 12\rangle \langle 34\rangle} \ .
\end{equation}
Since there are no triangle diagrams in the maximally supersymmetric theory, the BCJ relations follow from the fact that this coefficient is symmetric for permutations of the external particles. In the notation of eq.~(\ref{defnot}), we can capture the numerator above by
\begin{equation}
\num_{\underline{1}|2|3|4} = X_{2,4}X_{2,3} \  .
\label{box}
\end{equation}
The permutation symmetry follows from $X_{2,4}=-X_{2,3}=-X_{3,4}$, due to momentum conservation.
This symmetry is no longer manifest, but the benefit is that there is an extension to arbitrary multiplicity, i.e.~with tree-level subdiagrams $A_j$ in the corners of the box:
\beq
\num_{\underline{1}|A_2|A_3|A_4} = X_{A_2,A_4}X_{A_2,A_3}  \prod_{j=2}^4 X^{(A_j)} \ .
\label{massbox}
\eeq
The shorthand $X^{(A_j)}$ comprises a product of structure constants reflecting the vertices within the external tree-level subdiagram $A_j$ such as $X^{([2,3])}=X_{2,3}$ and $X^{([[2,3],4])}=X_{2,3}X_{2+3,4}$. More generally, any tree-level subdiagram can be reduced to the master topology of half-ladder trees
\beq
\tikzpicture[scale=0.6]
\draw(-8,0)node{$\displaystyle X^{([[\ldots [[2,3],4] \ldots,p-1],p])}= \prod_{j=2}^{p-1} X_{2+3+\ldots+j,j+1} \ \leftrightarrow$};
\draw(0,0)--(-1,-1)node[left]{2};
\draw(0,0)--(-1,1)node[left]{3};
\draw(1,0)--(1,1)node[above]{4};
\draw(2,0)--(2,1)node[above]{5};
\draw(2.75,0.5)node{$\ldots$};
\draw(3.5,0)--(3.5,1)node[above]{$p$};
\draw(0,0)--(4,0)node[right]{$\ldots$};
\endtikzpicture 
\label{moremass}
\eeq
by a sequence of kinematic Jacobi relations which in turn follow from the Schouten identity (\ref{Schout}). By symmetry of eq.~(\ref{massbox}) in $A_2,A_3,A_4$, the BCJ duality is consistent with the absence of bubbles and triangles \cite{Bern:1994zx,Bern:1994cg},
\beq
\num_{\underline{1}|A_2|A_3}  = \num_{\underline{1}|A_2}  = 0  \ .
\label{notriag}
\eeq
For $N=5$, the amplitude contains box and pentagon integrals, and the pentagons are the master topology in terms of the BCJ duality. The numerators of the pentagons and the boxes (with a massive corner) are, respectively,
\begin{align}
\num_{\underline{1}|2|3|4|5} & = X_{2,4}X_{2,3}X_{\ell,5}+X_{2,5}X_{2,3}X_{2+3,4}+X_{3,5}X_{\ell,2}X_{2+3,4} \  ,\nonumber \\
\num_{\underline{1}|2|3|[4,5]} & =   X_{2,4+5}X_{2,3}X_{4,5} =\num_{\underline{1}|2|[4,5]|3}=\num_{\underline{1}|[4,5]|2|3} \ .
\label{5ptnu}
\end{align}
The box numerator ties in with the massive generalization in eq.~(\ref{massbox}), with the factor of $X_{4,5}$ representing the tree-subdiagram as in eq.~(\ref{moremass}). One can easily check from eq.~(\ref{5ptnu}) that BCJ relations fix the boxes in terms of the pentagons,
\begin{equation}
\num_{\underline{1}|2|3|4|5} - \num_{\underline{1}|2|3|5|4} = \num_{\underline{1}|2|3|[4,5]} \ .
\end{equation}
The loop momentum dependence of the pentagon numerators cancels upon antisymmetrization since $X_{2,4}X_{2,3}X_{\ell,5}+X_{3,5}X_{\ell,2}X_{2+3,4}$ is permutation invariant in $2,3,4,5$. According to the general rule given below, an $m$-gon numerator ($4\leq m\leq N$) will be a polynomial of order $m-4$ in the loop momentum. Similar to the massive box in eq.~(\ref{massbox}), the massless pentagon numerator in eq.~(\ref{5ptnu}) generalizes as follows for massive corners:
\beq
\num_{\underline{1}|A_2|A_3|A_4|A_5} = (X_{A_2,A_4}X_{A_2,A_3}X_{\ell,A_5}+X_{A_2,A_5}X_{A_2,A_3}X_{A_2+A_3,A_4}+X_{A_3,A_5}X_{\ell,A_2}X_{A_2+A_3,A_4}) \prod_{j=2}^5 X^{(A_j)}  \ ,
\label{masspent}
\eeq
see eq.~(\ref{moremass}) for the contributions from tree-subdiagrams $A_j$.

For $N=6$, the amplitude contains box, pentagon and hexagon integrals. The hexagons furnish the master topology, and we additionally have massive pentagons as well as two types of massive boxes; see eqs.~(\ref{massbox}) and (\ref{masspent}). Explicitly,
\begin{align}
\num_{\underline{1}|2|3|4|5|6} &=   X_{2,4}X_{2,3}X_{\ell-6,5}X_{\ell,6}+X_{2,5}X_{2,3}X_{2+3,4}X_{\ell,6} +X_{2,6}X_{2,3}X_{2+3,4}X_{2+3+4,5}
\nonumber \\   & \qquad   +X_{3,5}X_{\ell,2}X_{2+3,4}X_{\ell,6} + X_{3,6}X_{\ell,2}X_{2+3,4}X_{2+3+4,5} + X_{4,6}X_{\ell,2}X_{\ell+2,3}X_{2+3+4,5} \ ,
\nonumber \\
\num_{\underline{1}|2|3|4|[5,6]} &=  \big(X_{2,4}X_{2,3}X_{\ell,5+6}+X_{2,5+6}X_{2,3}X_{2+3,4}+X_{3,5+6}X_{\ell,2}X_{2+3,4} \big)X_{5,6} \ ,
\label{hexnum} \\   
\num_{\underline{1}|2|[3,4]|[5,6]} &=   X_{2,5+6}X_{2,3+4}X_{3,4}X_{5,6} \ ,
\nonumber \\   
\num_{\underline{1}|2|3|[4,[5,6]]} &=  X_{2,4+5+6}X_{2,3}X_{4,5+6}X_{5,6} \ .
 \nonumber
\end{align}
The different types of BCJ relations relating the topologies are
\begin{align}
\num_{\underline{1}|2|3|4|5|6} - \num_{\underline{1}|2|3|4|6|5} &= \num_{\underline{1}|2|3|4|[5,6]} \ , \nonumber \\
\num_{\underline{1}|2|3|4|[5,6]} - \num_{\underline{1}|2|4|3|[5,6]} &= \num_{\underline{1}|2|[3,4]|[5,6]} \ ,  \\
\num_{\underline{1}|2|3|4|[5,6]} - \num_{\underline{1}|2|3|[5,6]|4} &= \num_{\underline{1}|2|3|[4,[5,6]]} \ . \nonumber
\end{align}
It will be shown in generality that the massive box and pentagon numerators in eqs.~(\ref{massbox}) and (\ref{masspent}) are compatible with the BCJ relations.

The resulting integrands (\ref{subamp}) for single-trace subamplitudes at $N=4$ and $N=5$ are given by
\begin{align}
\mathcal{I}_{1,2,3,4}(\ell)&=\frac{    X_{2,4} X_{2,3}}{\ell^2 (\ell+k_1)^2 (\ell+k_{12})^2 (\ell+k_{123})^2}
\label{4ptex} \\
\mathcal{I}_{1,2,3,4,5}(\ell)&=\frac{ X_{2,4}X_{2,3}X_{\ell,5}+X_{2,5}X_{2,3}X_{2+3,4}+X_{3,5}X_{\ell,2}X_{2+3,4} }{\ell^2 (\ell+k_1)^2 (\ell+k_{12})^2 (\ell+k_{123})^2 (\ell+k_{1234})^2}+ \frac{ X_{2,3} \ X_{2+3,4} X_{2+3,5}  }{s_{23} \ell^2 (\ell+k_1)^2  (\ell+k_{123})^2 (\ell+k_{1234})^2} \notag \\
 & + \frac{ X_{3,4} \  X_{2,3+4} X_{2,5}  }{s_{34}  \ell^2 (\ell+k_1)^2  (\ell+k_{12})^2 (\ell+k_{1234})^2}   + \frac{ X_{4,5} \ X_{2,3} X_{2,4+5}  }{s_{45}  \ell^2 (\ell+k_1)^2  (\ell+k_{12})^2 (\ell+k_{123})^2} 
  \ ,
\label{5ptex}
\end{align}
where the numerators $\num_{[1,2]|3|4|5}$ and $\num_{[5,1]|2|3|4}$ for two of the five massive box diagrams vanish thanks to $X_{1,2}=X_{5,1}=0$. The analogous six-point integrand is spelt out in appendix \ref{app:examples}. The remaining numerators beyond the canonically ordered single trace $1,2,\ldots,N$ can be obtained from eqs.~(\ref{4ptex}), (\ref{5ptex}) and (\ref{6ptex}) via permutations in $2,3,\ldots,N$. The resulting gravity amplitudes are discussed in subsection \ref{sec:sugra}.

The sets of BCJ numerators presented here differ from those in ref.~\cite{Carrasco:2011mn} for $N=5$ and in ref.~\cite{Bjerrum-Bohr:2013iza} for $N=6$. The inverse Gram determinants occurring in the numerators of those works do not arise in the string-theory setup, and the only non-localities in our numerators (\ref{defnot}) stem from polarization vectors, see eq.~(\ref{extstates}). As they are obtained from linear combinations of SYM tree-level subamplitudes, the numerators in this work provide closed-form expressions for any multiplicity.

\subsection{General result}
\label{jactypes}

Our formula for the BCJ numerators of one-loop MHV amplitudes is best understood by comparison to the one-loop all-plus numerators in eq.~\eqref{allplus}. We introduce a {\it prescription} $\mathscr{X}$ to represent this close relationship which enters the MHV $N$-gon numerator as follows:
\begin{equation}
n^{1-\text{loop,\;MHV}}_{\underline{1}|2|\cdots|N} = \delta^8(Q)\,  \left( \prod_{i=2}^N \frac{1}{\langle 1 i \rangle^2} \right)  \mathscr{X}_{\underline{1}}\left\{ \prod_{j=1}^N X_{\ell + 1+\cdots+(j-1),j} \right\} \ .
\end{equation}
The object acted upon by the operation $\mathscr{X}$ is simply the self-dual numerator (\ref{allplus}), up to universal prefactors. First, particle 1 was chosen by gauge choice to have a special role: only diagrams where particle 1 is directly attached to the loop are non-vanishing. Moreover, $\mathscr{X}_{\underline{1}}$ enforces $|\eta\rangle\to|1\rangle$. The non-trivial action of $\mathscr{X}$, however, is to eliminate four loop momenta from the numerator, such that the leading power is $\ell^{N-4}$ rather than $\ell^{N}$, as implied by maximal supersymmetry:
\begin{align}\label{gen_result}
&\mathscr{X}_{\underline{1}}\left\{ \prod_{j=1}^N X_{\ell + 1+\cdots+(j-1),j} \right\} = 
\sum_{1<r<s-1}^N
X_{r,s} \;\; \times \nonumber \\  & \qquad \times\;\; 
\left(\prod_{i=2}^{r-1} X_{\ell+1+\cdots+(i-1),i}\right)
\left(\prod_{j=r+1}^{s-1} X_{1+\cdots+(j-1),j}\right)
\left(\prod_{k=s+1}^{N} X_{\ell+1+\cdots+(k-1),k}\right) \ .
\end{align}
Let us break down the prescription $\mathscr{X}$:
\begin{itemize}
\item First, all the $X$ on the right-hand-side are defined as in eq.~(\ref{defXX}) with $|\eta\rangle=|1\rangle$.
\item Take the $N$-gon and give it a boundary set by particle 1, deleting the vanishing factor $X_{\ell,1}$.
\item Each pair of non-neighbouring corners of the $N$-gon, labelled by $r$ and $s$ ($s>r+1$), produces a contribution given by $X_{r,s}$ times a product of the $X$ factors excluding $X_{\cdots,r}$ and $X_{\cdots,s}$.
\item The lowering of the degree in the loop momentum is achieved by deleting $\ell$ from the $X$'s representing vertices located between $r$ and $s$.
\end{itemize}
Going back to the examples presented above, at four points,
\begin{equation}
\mathscr{X}_{\underline{1}}\left\{ X_{\ell,1}X_{\ell+1,2}X_{\ell+1+2,3}X_{\ell,4} \right\} = 
X_{2,4} X_{2,3} \ ,
\end{equation}
while at five points,
\begin{equation}
\mathscr{X}_{\underline{1}}\left\{ X_{\ell,1}X_{\ell+1,2}X_{\ell+1+2,3}X_{\ell-5,4}X_{\ell,5} \right\} = 
X_{2,5} X_{2,3} X_{4,5} + X_{2,4} X_{2,3} X_{\ell,5} + X_{3,5} X_{\ell,2} X_{4,5} \ ,
\end{equation}
reproducing the expressions in eqs.~(\ref{box}) and (\ref{5ptnu}).

Since the prescription $\mathscr{X}$ does not depend on the corners being massless, it can be applied directly not only to the $N$-gon but also to any massive $m$-gon numerator with $4\leq m\leq N$,
\begin{align}\label{massgen}
&\num_{\underline{1}|A_2|A_3|\ldots |A_m}= 
\left(\prod_{p=2}^m X^{(A_p)}  \right)
\sum_{1<r<s-1}^m
X_{A_r,A_s} \;\; \times \nonumber \\  & \qquad \times\;\; 
\left(\prod_{i=2}^{r-1} X_{\ell+A_2+\cdots+A_{i-1},A_i}\right)
\left(\prod_{j=r+1}^{s-1} X_{A_2+\cdots+A_{j-1},A_j}\right)
\left(\prod_{k=s+1}^{m} X_{\ell+A_2+\cdots+A_{k-1},A_k}\right) \ ;
\end{align}
see eq.~(\ref{moremass}) for the contribution $X^{(A_p)} $ of the tree-level subdiagram $A_p$. Recall that the only non-vanishing numerators are the ones with particle 1 attached directly to the loop. Up to multiplicity $N=10$, the string-theory origin of the numerators in eq.~(\ref{massgen}) has been worked out explicitly, and their validity at higher $N$ is a conjecture supported by their connection with all-plus amplitudes.

In order to verify that the MHV numerators constructed in this way satisfy the BCJ duality, three different classes of kinematic Jacobi relations need to be considered:
\begin{itemize}
\item Jacobi identities affecting the propagators of tree-level subdiagrams hold by the Schouten identity (\ref{Schout}) among the structure constants in eq.~(\ref{moremass}).
\item Jacobi identities affecting the $m$-gon propagators non-adjacent to leg 1 hold by the property
\beq
\num_{\underline{1}|A_2|A_3|\ldots |A_i|A_{i+1}| \ldots |A_m}  - \num_{\underline{1}|A_2|A_3|\ldots |A_{i+1}|A_{i}| \ldots |A_m}  = \num_{\underline{1}|A_2|A_3|\ldots |[A_i,A_{i+1}]| \ldots |A_m} 
\label{easyjac}
\eeq
of the general numerator in eq.~(\ref{massgen}). The large tree subdiagram $[A_i,A_{i+1}]$ on the right-hand side comprises the smaller tree subdiagrams $A_i$ and $A_{i+1}$ connected through a cubic vertex; accordingly it contributes the structure constants $X^{([A_i,A_{i+1}])}= X_{A_i,A_{i+1}}X^{(A_i)}X^{(A_{i+1})}$, see eq.~(\ref{moremass}). As a proof of eq.~(\ref{easyjac}), we will demonstrate in appendix \ref{proofBCJ} that the prescription $\mathscr{X}$ commutes with this class of Jacobi identities. This remarkable fact is easy to check for certain choices of pairs $\{r,s\}$, and we have already seen the simplest examples at $N=4,5,6$. 
\item Jacobi identities affecting the $m$-gon propagators adjacent to leg 1 hold by the property
\beq
\num_{\underline{1}|A_2|A_3|\ldots |A_m}(\ell) = \num_{\underline{1}|A_3|A_4|\ldots |A_m|A_2}(\ell+k_{A_2}) = \num_{\underline{1}|A_m|A_2|A_3|\ldots |A_{m-1}}(\ell-k_{A_m}) 
\label{hardjac}
\eeq
of the general numerator in eq.~(\ref{massgen}). It is crucial to use the same conventions for loop momenta in the three graphs related by a Jacobi identity. Since the momentum $\ell$ in the argument of the numerator $\num_{\underline{1}|A_2|A_3|\ldots |A_m}(\ell)$ is defined to reside in the $m$-gon edge prior to leg 1, the antisymmetrization in corners $1,A_2$ or $1,A_m$ leads to a shift as $\ell \mapsto \ell+k_{A_2}$ or $\ell \mapsto \ell-k_{A_m}$, respectively. The would-be third diagram in this class of Jacobi identities is an $(m{-}1)$-gon with massive corner $[1,A_2]$ or $[1,A_m]$ whose numerator vanishes by construction. While eq.~(\ref{hardjac}) is straightforward to check analytically at low multiplicity $m$, we do not give a general proof and rely on numerical checks. We have checked that eq.~(\ref{hardjac}) holds for $m\leq 20$. 
\end{itemize}

\subsection{Relation to dimension-shifting formula}

There is a well-known relation at one loop between the integrand of MHV amplitudes in the maximally supersymmetric gauge theory and the (integrated) all-plus amplitudes of the non-supersymmetric theory; there is also a gravity counterpart. This is the dimension-shifting formula of ref.~\cite{Bern:1996ja}. It can be expressed as
\begin{equation}
\label{eq:SDMHV}
\delta^8(Q)\, A_{\text{all-plus}}^{1-\text{loop}} = -2 \epsilon (1-\epsilon) (4\pi)^2 \,
A_{\text{MHV}}^{1-\text{loop}}  \Big|_{D\to D+4}\ ,
\end{equation}
where $D=4-2\epsilon$, and the delta function on the left hand side compensates for the different amount of supersymmetry preserved by $A_{\text{all-plus}}^{1-\text{loop}}$ and $A_{\text{MHV}}^{1-\text{loop}}$, respectively. The dimension shift $D\to D+4$ of the MHV amplitude is to be taken before integration, when the amplitude is expressed in terms of $D$-dimensional scalar integrals. In the MHV integrand, the only surviving diagrams after the shift in the limit $\epsilon\to0$ are the box integrals. 

\begin{figure}[t]
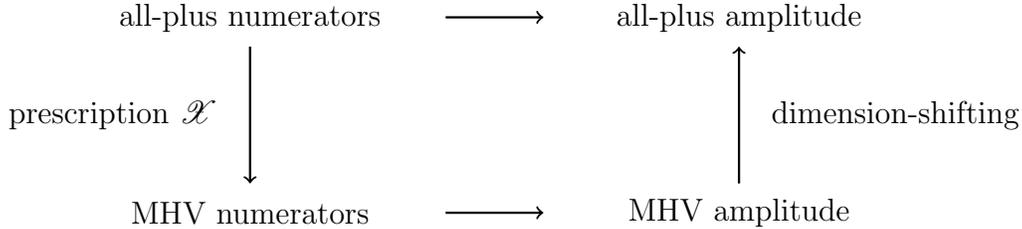

\begin{center}
 \tikzpicture [scale=1.3]
 \draw (0,0) node {MHV numerators};
 \draw (0,2) node {all-plus numerators};
 \draw (-1.4,1) node {prescription $\mathscr{X}$};
 \draw [line width=0.30mm,->] (0,1.7) -- (0,0.3);
 \draw (5,0) node { MHV amplitude };
 \draw (5,2) node { all-plus amplitude };
 \draw (6.6,1) node {dimension-shifting};
 \draw [line width=0.30mm,<-] (5,1.7) -- (5,0.3);
 \draw [line width=0.30mm,->] (2,0) -- (3,0);
 \draw [line width=0.30mm,->] (2,2) -- (3,2);
 \endtikzpicture
\end{center}
\caption{Our prescription $\mathscr{X}$ versus the dimension-shifting formula of ref.~\cite{Bern:1996ja}.}
\label{figflow}
\end{figure}

We saw above that a sort of converse statement is true: the BCJ numerators of all-plus amplitudes allow us to write down the BCJ numerators of MHV amplitudes. We can summarize the relationships as in figure \ref{figflow}.

\subsection{Supergravity amplitudes}
\label{sec:sugra}
We can obtain one-loop MHV amplitudes in ${\cal N}=8$ supergravity through the double-copy construction by squaring the numerators of ${\cal N}=4$ super-Yang-Mills (SYM),
\beq
M^{1-\te{loop}}_{N,\te {MHV}}=\frac{\delta^{16}(Q)}{\prod_{i=2}^N \langle 1 i\rangle^4}
\int \frac{\dd^D\ell}{(2\pi)^D}  \sum_{m=4}^N  \sum_{ A_2\cup A_3 \cup  \ldots \cup A_m \atop{= \{2,3,\ldots,N\}}} \frac{\num_{\underline{1}| A_2| \ldots | A_m}^2(\ell) 
}{\prod_k P_{k;A_2,\ldots, A_m} (\ell)} \ ,
\label{sugra1}
\eeq
where $\num_{\underline{1} | A_2| \ldots | A_m}(\ell)$ is given by eq.~(\ref{massgen}); 
the sum is over all possible one-loop $m$-gon diagrams with sub-trees $A_2,\ldots A_m$ attached to the corners, and the denominator is the product of all propagators of the diagram. Note that the $N=5$ instance of eq.~(\ref{sugra1}) descends from the ten-dimensional supergravity amplitude in ref.~\cite{Mafra:2014gja} upon dimensional reduction and insertion of helicity wavefunctions.

As discussed in details in ref.~\cite{Bern:2011rj} (see eqs.~(3.4) and (3.6) therein), an alternative form of the one-loop double-copy is given in terms of master $N$-gon numerators and full gauge-theory integrands,
\beq
M^{1-\te{loop}}_{N} = \int \frac{\dd^D\ell}{(2\pi)^D} \sum_{\sigma \in S_{N-1}} n^{1-\te{loop}}_{1|\sigma(2)|\sigma(3)|\ldots|\sigma(N)} (\ell)\,\tilde I_{1,\sigma(2),\sigma(3),\ldots,\sigma(N)}^{1-\te{loop}}(\ell)\ ,
\eeq
where the sum is over all cyclically inequivalent permutations $\sigma$, and $\tilde I^{1-\te{loop}}$ may in general be associated with numerators for different states, similarly to the numerators $\tilde n$ in eq.~\eqref{LoopSq}. For MHV in the maximally supersymmetric theories, we can use explicit formulae for both the numerators, (\ref{gen_result}), and the full integrand (see e.g. eqs.~(\ref{4ptex}), (\ref{5ptex}) and (\ref{6ptex})), and obtain
\beq\label{sugra}
M^{1-\te{loop}}_{N,\te {MHV}} = \frac{\delta^{16}(Q)}{\prod_{i=2}^N \langle 1 i\rangle^4} \int \frac{\dd^D\ell}{(2\pi)^D} \sum_{\sigma \in S_{N-1}} \num_{\underline{1}|\sigma(2)|\sigma(3)|\ldots|\sigma(N)} (\ell)\,\mathcal{I}_{1,\sigma(2),\sigma(3),\ldots,\sigma(N)}(\ell)\ .
\eeq
As shown in ref.~\cite{Bern:2011rj}, an important feature of this representation is that the gauge-theory integrands $\tilde I^{1-\te{loop}}$ do not have to be in BCJ form to give correct supergravity result. Thus, given any representation for the SYM integrand (or equivalently the stripped version $\mathcal{I}$), (\ref{sugra}) and our result (\ref{gen_result}) directly yield a formula for MHV integrands of ${\cal N}=8$ supergravity. More importantly, one can combine the numerators of ${\cal N}=4$ SYM as in eq.~(\ref{massgen}) with integrands $I_{1,\sigma(2),\sigma(3),\ldots,\sigma(N)}^{1-\te{loop}}(\ell)$ of less supersymmetric ${\cal N}<4$ gauge theories to obtain one-loop amplitudes of $(4+{\cal N})$ supergravities \cite{Bern:2011rj}.

\section{String-theory derivation at one-loop}\label{sec5}
\label{sec:string}

This section is devoted to a derivation of the BCJ numerators in eq.~(\ref{massgen}) from superstring theory.

\subsection{Schwinger parametrization of Feynman integrals}

The appearance of Feynman integrals in the $\ap \rightarrow 0$ limit of superstring amplitudes is particularly transparent in their Schwinger parametrization. For a scalar $m$-gon integral in $D$ dimensions, this amounts to a change of variables \cite{Bern:1990cu, Bern:1991aq, Strassler:1992zr, BjerrumBohr:2008ji},
\begin{align}
&\int \frac{  \dd^D \ell  }{ \ell^2 \, (\ell+k_1)^2 \, (\ell+k_{12})^2 \, \ldots \, (\ell + k_{12\ldots m-1})^2 } = \pi^{m} \int^{\infty}_0 \frac{ \dd t}{t} \, t^{m-D/2 }  \! \! \! \! \! \! \int \limits_{0 \leq \nu_i \leq \nu_{i+1} \leq 1} \! \! \! \! \! \! \dd \nu_2  \, \ldots \, \dd \nu_m \, e^{ - \pi t  Q_m}  \ ,
\label{loopnum1}
\end{align}
towards worldline length $t$ and proper times $\nu_j$ with the following shorthand in the exponent
\beq
Q_m \equiv \sum_{i<j}^m(k_i \cdot k_j) \, \Big( \nu_{ij}^2 \, - \, |\nu_{ij}| \Big)  \co \nu_1=0 \co \nu_{ij} \equiv \nu_i - \nu_j\ .
\label{loopnum2}
\eeq
For tensorial Feynman integrals, i.e.~in presence of loop momenta $T_{\mu_1\mu_2\ldots \mu_r} \ell^{\mu_1} \ell^{\mu_2} \ldots \ell^{\mu_r}$ in the numerator, the analogous expressions take a particularly simple form if the accompanying symmetric tensor is traceless, $\eta^{\mu_1 \mu_2}T_{\mu_1\mu_2\ldots \mu_r}=0 $:
\begin{align}
&\int \frac{  \dd^D \ell \ T_{\mu_1\mu_2\ldots \mu_r} \ell^{\mu_1} \ell^{\mu_2} \ldots \ell^{\mu_r}  }{ \ell^2 \, (\ell+k_1)^2 \, (\ell+ k_{12})^2 \, \ldots \, (\ell + k_{12\ldots m-1})^2 } \Big|_{\te{traceless}}= \pi^{m} \int^{\infty}_0 \frac{ \dd t}{t} \, t^{m-D/2 } \notag \\
 & \ \ \ \ \ \ \ \ \ \ \ \ \ \ \ \ \times  \! \! \! \! \! \! \int \limits_{0 \leq \nu_i \leq \nu_{i+1} \leq 1} \! \! \! \! \! \! \dd \nu_2  \, \ldots \, \dd \nu_m \, e^{ - \pi t  Q_m}  \, T_{\mu_1\mu_2\ldots \mu_r} L^{\mu_1} L^{\mu_2} \ldots L^{\mu_r}  \ .
\label{loopnum3}
\end{align}
The right hand side involves the following shift of loop momentum
\beq
L^\mu \equiv \sum_{i=2}^m k^\mu_i \, \nu_i  \ .
\label{loopnum4}
\eeq
Since any loop momentum in the BCJ numerators (\ref{massgen}) enters through the kinematic structure constants, the requirement of having traceless tensors in eq.~(\ref{loopnum3}) is automatically satisfied\footnote{In four-dimensional helicity configurations beyond MHV and in the ten-dimensional six-point amplitude of ref.~\cite{Mafra:2014gja}, tensorial numerators are no longer traceless, and eq.~(\ref{loopnum3}) receives extra terms with $L^{\mu_i} L^{\mu_j} \rightarrow \eta^{\mu_i \mu_j} /( 2\pi t)$.}:
\beq
X_{\ell,i} X_{\ell,j} = \ell^\mu \ell^\nu \langle 1 | \sigma_\mu k_i |1 \rangle  \langle 1 | \sigma_\nu k_j | 1\rangle  \co \eta^{\mu \nu} \langle 1 | \sigma_\mu k_i |1 \rangle  \langle 1 | \sigma_\nu k_j |1 \rangle  = 0 \ .
\label{loopnum5}
\eeq
Hence, eq.~(\ref{loopnum3}) translates the $r$'th power of loop momentum in the numerators (\ref{massgen}) to a polynomial of degree $r$ in the proper times $\nu_j$. Note that eq.~(\ref{loopnum3}) does not depend on the external masses and remains valid for off-shell momenta $k_j^2 \neq 0$ in case of external tree-level subdiagrams.

\subsection{Feynman integrals from string theory}

We will next illustrate how the worldline integrals on the right hand side of eq.~(\ref{loopnum3}) arise in the field-theory limit of one-loop superstring amplitudes. For the open string, the worldline length $t$ descends from the modular parameter $\tau$ of the cylindrical worldsheet\footnote{The string amplitude also receives contributions from worldsheets of M\"obius strip topology and ``non-planar'' cylinders with state insertions on both boundaries \cite{Green:1987mn}. Since the planar cylinder already yields unique answers for the BCJ numerators, we will neglect the additional worldsheet topologies in the sequel. From the string-theory perspective, however, their interplay is crucial for the cancellation of infinities \cite{Green:1984ed} and anomalies \cite{Green:1984sg, Green:1984qs} for gauge group $SO(32)$.}, and the ordering of the proper times $\nu_i \leq \nu_{i+1}$ is inherited from the arrangement of vertex operators on the cylinder boundary. In this setting, single-trace one-loop amplitudes among massless open superstring excitations in $D$ spacetime dimensions take the schematic form
\begin{align}
{\cal A}^{1-\te{loop}}(1,2,\ldots,N) =
\int^{\infty}_0 \frac{ \dd \tau }{\tau^5} \, \Ga_{10-D}(\tau) 
\! \! \! \! \! \! \! \! \! \! \! \! \!  
 \int \limits_{0 \leq  \Im(z_{i}) \leq \Im(z_{i+1 })  \leq \tau} 
 \! \! \! \! \! \! \! \! \! \! \! \! \!  
  \dd z_2 \,\ldots \, \dd z_{N}  \,e^{-  2{\cal Q}_N} \, {\cal K}_N
   \ .
   \label{loopnum6}
  \end{align}
The lattice factor $\Ga_{10-D}(\tau) $ describes supersymmetry-preserving compactifications of ten-dimensional Minkowski spacetime on $10-D$ dimensional tori; their field-theory limit yields maximally supersymmetric Yang-Mills theories in $D<10$ dimensions \cite{Green:1982sw}. The cylinder punctures $z_j$ with $\Re(z_j)=0$
enter through the correlation function of $N$ vertex operators on a genus-one Riemann surface, where the one-loop instance $e^{-  2{\cal Q}_N} $ of the Koba-Nielsen factor $\prod_{i<j}|z_i-z_j|^{\alpha' k_i\cdot k_j}$ in eq.~(\ref{treenum1}) can be factored out and reads
\beq
\label{KNdef}
{\cal Q}_N \equiv   \sum_{i<j}^N k_i \cdot k_j \, {\cal G}(z_i,z_j,\tau)  \co {\cal G}(z_i,z_j,\tau) \equiv -  \frac{\ap}{2} \, \ln \big| \, \theta_1(z_{i} - z_{j},i\tau) \, \big|^2
+ \frac{\ap \pi}{\tau} \, [\Im (z_{i}-z_{j})]^2 \ .
\eeq
The bosonic Green function ${\cal G}(z_i,z_j,\tau)$ on a genus-one surface involves the odd Jacobi theta function $\theta_1$. The external polarizations enter through the remaining integrand ${\cal K}_N$ in eq.~(\ref{loopnum6}) whose representation depends on the formalism chosen. The systematic evaluation and economic presentation of ${\cal K}_N$ pose one major challenge in the analysis of multiparticle one-loop amplitudes of the open superstring, see e.g.~refs.~\cite{Tsuchiya:1988va, Montag:1992dm, Stieberger:2002wk, BjerrumBohr:2008vc, Mafra:2009wi, Mafra:2012kh}. The simplest nonvanishing instances are
\begin{align}
{\cal K}_4 &= s_{23} s_{34} A^{\te{tree}}(1,2,3,4)
\co
{\cal K}_5 = \partial {\cal G}(z_2,z_3,\tau) s_{23} C_{1|23,4,5} + (2,3\leftrightarrow 2,3,4,5) \ ,
\label{loopnum7}
\end{align}
with five-point kinematic factors\footnote{In refs.~\cite{Mafra:2012kh, Mafra:2014oia, Mafra:2014gsa}, the notation $C_{1|A,B,C}$ refers to BRST-invariant expressions in pure-spinor superspace which enter kinematic factors through the zero mode bracket $\langle C_{1|A,B,C} \rangle$ explained below eq.~(\ref{treenum2}). In order to avoid the ubiquitous appearance of $\langle \ldots \rangle$, this operation is absorbed into this work's definition of $C_{1|A,B,C}$. Also, the conventions for Mandelstam invariants in eq.~(\ref{mand}) differ from these references and give rise to conversion factors of 2.}
\beq
C_{1|23,4,5} \equiv s_{45} \, \big( s_{24}  A^{\te{tree}}(1,3,2,4,5) \, - \, s_{34}  A^{\te{tree}}(1,2,3,4,5)  \big)
\label{loopnum8}
\eeq
and partial SYM tree-level amplitudes $A^{\te{tree}}(1,2,\ldots,N)$. Higher-multiplicity expressions obtained from the pure-spinor formalism in ref.~\cite{Mafra:2012kh} will be reviewed in the subsequent subsection.

In order to recover Feynman integrals from the string prescription (\ref{loopnum6}), the point-particle limit $\ap\to 0$ must be accompanied by a degeneration of the cylindrical worldsheet to a worldline diagram via $\tau \to\infty$. Moreover, the combined field-theory limit must be performed such that the proper time $t \equiv \ap \tau$ and the 
proper times $\nu_j\equiv \frac{ \Im(z_j) }{\tau}$ stay finite. The worldsheet Green function in eq.~(\ref{KNdef}) then reduces to the worldline Green function $G_{ij}$,
\beq
{\cal G}(z_i,z_j,\tau) \rightarrow  \frac{ \pi t}{2} \, \big( \nu_{ij}^2 \ - \ |\nu_{ij}| \big) \equiv G_{ij} \co
\pa {\cal G}(z_i,z_j,\tau) \rightarrow   \pi \, \big(\nu_{ij}  - \tfrac{1}{2} \, \te{sgn}_{ij}\big) \equiv \partial G_{ij} \ ,\label{loop3d}
\eeq
where $\te{sgn}_{ij}\equiv\te{sgn}(\nu_{ij})$ is defined to be $+1 (-1)$ when $\nu_i\ge\nu_j$ ($\nu_i<\nu_j$). As a consequence, the Koba--Nielsen factor (\ref{KNdef}) behaves as
\beq
\label{KNFT}
e^{-2 {\cal Q}_N} \rightarrow  e^{ - \pi t \, Q_N  } \  ,
\eeq
reproducing the universal integrand (\ref{loopnum3}) in the Schwinger parametrization of Feynman integrals.

In compactifications of ten-dimensional Minkowski spacetime on a $10-D$ dimensional torus, Kaluza-Klein and winding modes decouple when the radii $R$ of the toroidal directions and the string length $\sqrt{\ap}$ vanish according to the following scaling limit \cite{Green:1982sw}:
\beq
  R\, \rightarrow \, 0 \co \frac{\ap}{ R }\,\rightarrow \,0 \ \ \ \Rightarrow \ \ \  
 \Ga_{10-D}(\tau)  \rightarrow (R \tau)^{5-D/2}
\ .
\label{loop3e}
\eeq
The resulting scaling of the lattice factor $\Ga_{10-D}$ keeps the $D$-dimensional SYM coupling finite \cite{Green:1982sw}.
Therefore, the field-theory limit $\ap\to 0$ and $\tau \to\infty$ maps the string-theory measure (\ref{loopnum6}) to the Schwinger representation of Feynman integrals as in eq.~(\ref{loopnum3}),
\begin{align}
&\int^{\infty}_0 \frac{ \dd \tau }{\tau^5} \, \Ga_{10-D}(\tau) 
\! \! \! \! \! \! \! \! \! \! \! \! \! 
 \int \limits_{0 \leq  \Im(z_{i}) \leq \Im(z_{i+1 })  \leq \tau} 
 \! \! \! \! \! \! \! \! \! \! \! \! \! 
  \dd z_2 \,\ldots \, \dd z_{N} \, e^{ -  2 {\cal Q}_N} \, {\cal K}_N \label{loopnum9} \\
  &\ \ \ \ \rightarrow \   \int^{\infty}_0 \frac{ \dd t}{t} \, t^{N-D/2 } \ \! \! \! \! \! \! \int \limits_{0 \leq \nu_i \leq \nu_{i+1} \leq 1} \! \! \! \! \! \! \dd \nu_2  \, \ldots \, \dd \nu_N \, e^{ - \pi t  Q_N}  \, K_N  + {\cal O}(s_{i_1\ldots i_p}^{-1}) \ ,  \notag
\end{align}
see ref.~\cite{Tourkine:2013rda} for an extension to higher loops. This degeneration prescription does not yet take kinematic poles ${\cal O}(s_{i_1\ldots i_p}^{-1})$ into account; these originate from singularities of the worldsheet integrand ${\cal K}_N$ as $z_i \rightarrow z_j$. They are associated with $m<N$-gon integrals in an $N$-point amplitude and will be discussed separately in subsection \ref{sec:poles}. The $N$-gon numerator of the SYM amplitude, i.e.~its irreducible piece, can be reliably extracted from eq.~(\ref{loopnum9}) by translating the Green functions in the integrands as in eq.~(\ref{loop3d}),
\beq
K_N  \equiv {\cal K}_N  \Big|_{
\pa {\cal G}(z_i,z_j,\tau) \rightarrow  \partial G_{ij} 
} \ .
\label{loopnum10}
\eeq
This defines the worldline counterpart $K_N$ of the string-theory integrand ${\cal K}_N$. Since the superstring integrands ${\cal K}_N$ of interest to this work will be polynomials in $\partial {\cal G}$ of degree $N-4$, see e.g.~eq.~(\ref{loopnum7}) at $N=4,5$, the worldline integrands $K_N$ in eq.~(\ref{loopnum10}) become polynomials in proper times $\nu_j$ by (\ref{loop3d}). Hence, they have the right structure to identify the field-theory limit (\ref{loopnum9}) with tensorial $N$-gon integrals in their Schwinger parametrization (\ref{loopnum3}).

\subsection{The non-anomalous one-loop correlator}

In this subsection, we describe the all-multiplicity structure of the open string integrand ${\cal K}_N$ in eq.~(\ref{loopnum6}). The singular factors of $ {\cal G}(z_i,z_j,\tau) \sim (z_i-z_j)^{-1}$ in its five-point instance (\ref{loopnum7}) reflect one OPE among the vertex operators, and we have used integration by parts relations
\begin{align}
0 &= -  \int \dd z_2 \, \frac{\dd }{\dd z_2} \,  e^{-  2 {\cal Q}_5} \label{loopnum11} \\
&=\int \dd z_2 \, e^{- 2 {\cal Q}_5}  \, \big[ s_{23} \partial {\cal G}(z_2,z_3,\tau)
+s_{24} \partial {\cal G}(z_2,z_4,\tau)+s_{25} \partial {\cal G}(z_2,z_5,\tau)-s_{12} \partial {\cal G}(z_1,z_2,\tau)
\big]
\notag 
\end{align}
to eliminate the four instances of $ {\cal G}(z_1,z_j,\tau)$ with $j=2,3,4,5$. The resulting six basis integrals over ${\cal G}(z_i,z_j,\tau)$ with $2 \leq i<j\leq 5$ are accompanied by BRST-invariant kinematic factors (\ref{loopnum8}) in pure-spinor superspace. This approach has been extended to higher multiplicity in ref.~\cite{Mafra:2012kh}: Polynomials in ${\cal G}(z_i,z_j,\tau)$ which are independent under integration by parts have been matched with BRST-invariant quantities $C_{1|A,B,C}$ built from iterated OPEs among vertex operators. The shorthand
\beq
Y_{ij} \equiv s_{ij} \partial {\cal G}(z_i,z_j,\tau) \,
\label{loopnum12}
\eeq
simplifies both the integration-by-parts relations (\ref{loopnum11}) and the higher-multiplicity integrands~${\cal K}_N$
\begin{align}
{\cal K}_6 &= \big[Y_{23}(Y_{24}+ Y_{34}) C_{1|234,5,6} + Y_{24}(Y_{23}+ Y_{43}) C_{1|243,5,6}
+ (234\leftrightarrow 235,236,\ldots,456) \big]\notag \\
& \ \ + \big[ Y_{23} Y_{45} C_{1|23,45,6}+ Y_{24} Y_{35} C_{1|24,35,6}+ Y_{25} Y_{34} C_{1|25,34,6} + (6 \leftrightarrow 5,4,3,2) \big]\ ,
\label{loopnum13} \\
{\cal K}_{7} &= 15 \ \te{terms} \ \big[Y_{23}(Y_{24}+ Y_{34})(Y_{25}+ Y_{35} + Y_{45}) C_{1|2345,6,7} + \te{perm}(3,4,5) \big]\notag \\
& \ \ +  \ 60 \ \te{terms} \ \big[ Y_{23}(Y_{24}+ Y_{34}) C_{1|234,56,7} + Y_{24}(Y_{23}+ Y_{43}) C_{1|243,56,7} \big] Y_{56} \notag \\
& \ \ + \ 15 \ \te{terms} \ \big[ Y_{23} Y_{45} Y_{67} C_{1|23,45,67}  \big] \ .
\label{loopnum14}
\end{align}
For ${\cal K}_{7}$ we simply indicate the number of terms of a given type, and the all-multiplicity pattern can be found in subsection 5.4 of ref.~\cite{Mafra:2012kh}. Remarkably, the BRST-invariant quantities $C_{1|A,B,C}$ can be expressed in terms of tree-level subamplitudes of ten-dimensional SYM \cite{Mafra:2012kh}, as in eq.~(\ref{loopnum8}) and
\begin{align}
 C_{1|234,5,6}  & =
       s_{56} \big[ s_{45} A^{\rm tree}(1,2,3,4,5,6)   
       + s_{25} A^{\rm tree}(1,4,3,2,5,6)\notag \\
       & \ \ \ \ \ \ - s_{35} (A^{\rm tree}(1,2,4,3,5,6)+ A^{\rm tree}(1,4,2,3,5,6))  \big] \notag \\
 C_{1|23,45,6}  & =
        s_{46} s_{36} A^{\rm tree}(1,2,3,6,4,5)
       - s_{56} s_{36} A^{\rm tree}(1,2,3,6,5,4)\label{loopnum15} \\
& \ \ \ \ \ \      - s_{46} s_{26} A^{\rm tree}(1,3,2,6,4,5)
       + s_{56} s_{26} A^{\rm tree}(1,3,2,6,5,4)\ ,\notag
\end{align}
see appendix B of ref.~\cite{Mafra:2014oia} for the general pattern. When reducing to the MHV case in $D=4$, the Parke-Taylor formula \cite{Parke:1986gb} allows us to obtain compact spinor-helicity expressions, e.g.
\begin{align}
C_{1|23,4,5} \, \Big|_{\te{MHV}}&= \frac{ - \delta^8(Q) [45]^2 }{\langle 12 \rangle \, \langle 23 \rangle \, \langle 31 \rangle }
\notag\\
C_{1|234,5,6} \, \Big|_{\te{MHV}}&= \frac{ - \delta^8(Q)  [56]^2 }{\langle 12 \rangle \, \langle 23 \rangle \, \langle 34 \rangle \, \langle 41 \rangle }
\label{loopnum16} \\
C_{1|23,45,6} \, \Big|_{\te{MHV}}&= \frac{ \delta^8(Q)  \langle 1|\,( k_2 + k_3) \, |6]^2 }{\langle 12 \rangle \, \langle 23 \rangle \, \langle 31 \rangle \, \times \, \langle 14 \rangle \, \langle 45 \rangle \, \langle 51 \rangle }
\notag
\end{align}
with the following generalization:
\begin{align}
&C_{1|23\ldots p , p+1 \ldots q , q+1 \ldots n} \, \Big|_{\te{MHV}} =
\frac{- \delta^8(Q)  \, \langle 1| \, (k_2+k_3 + \ldots +k_p) \, (k_{p+1}+ \ldots +k_q ) \, |1 \rangle^2 }{\langle 12 \rangle \, \langle 23 \rangle \, \ldots \, \langle p1 \rangle  } \notag \\
& \ \ \ \ \ \ \times \, \frac{1}{ \langle 1, p+1 \rangle \, \langle p+1,p+2 \rangle \, \ldots \, \langle q1 \rangle \, \times \, \langle 1, q+1 \rangle \, \langle q+1,q+2 \rangle \, \ldots \, \langle n1 \rangle} \ .
\label{loopnum17}
\end{align}
We should emphasize that, due to their origin from pure-spinor superspace, the kinematic factors $C_{1|A,B,C}$ by themselves comprise {\em ten}-dimensional SYM amplitudes, so the expressions in eqs.~(\ref{loopnum16}) and (\ref{loopnum17}) are special cases for lower dimensions and MHV helicity configurations.

However, there is a major shortcoming to the $(N\geq 6)$-point integrands ${\cal K}_N$ obtained in ref.~\cite{Mafra:2012kh} by imposing BRST invariance: The open superstring is plagued by a hexagon gauge anomaly unless the gauge group is chosen as $SO(32)$ \cite{Green:1984sg, Green:1984qs}. The anomaly cancellation relies on the interplay between different worldsheet topologies, so the full integrand cannot be gauge invariant for $N\geq 6$. In the pure-spinor formalism, the resulting BRST anomaly at $N=6$ has been determined in ref.~\cite{Berkovits:2006bk}. The integrand of the underlying anomalous amplitudes must comprise an extension of the BRST invariant ${\cal K}_N$ in ref.~\cite{Mafra:2012kh}, the expressions in eqs.~(\ref{loopnum13}) and (\ref{loopnum14}) cannot capture the complete worldsheet correlation function at $N=6$ and $N=7$. 

Kinematic factors which carry the fingerprints of the BRST anomaly have been constructed in ref.~\cite{Mafra:2014gsa}. Their appearance in the six-point SYM amplitude at one loop has been described in ref.~\cite{Mafra:2014gja}, and their embedding into full-fledged superstring integrands will be explained in upcoming work\footnote{In the language of ref.~\cite{Broedel:2014vla}, the anomalous part of the correlator is captured by worldsheet functions $f^{(2)},f^{(3)},\ldots$ of higher modular weight which for instance arise from the failure of $\partial {\cal G} \equiv-\frac{\alpha'}{2} f^{(1)}$ to satisfy partial fraction relations,
\[
f^{(1)}(z_1-z_2,\tau) f^{(1)}(z_1-z_3,\tau)+ \te{cyc}(1,2,3)=
 f^{(2)}(z_1-z_2,\tau)+ \te{cyc}(1,2,3) \ .
\]}. Six-point and seven-point instances of these anomalous kinematic factors turn out to vanish for MHV helicity configurations\footnote{We would like to thank Carlos Mafra for contributing to these checks.} upon dimensional reduction to $D=4$. Even though the all-multiplicity pattern requires further investigation, this supports the idea that the non-anomalous integrands of ref.~\cite{Mafra:2012kh} including eqs.~(\ref{loopnum13}) and (\ref{loopnum14}) give correct MHV amplitudes. As will be detailed in the sequel, the matching of their worldline limit (\ref{loopnum10}) with Feynman integrals (\ref{loopnum3}) provides very strong consistency checks, which have been established up to multiplicity $N=10$. 

\subsection{Irreducible pieces}
\label{sec:irred}

In this subsection, we illustrate the procedure to construct the $N$-gon numerator (\ref{gen_result}) of $N$-point MHV amplitudes from the above string integrands ${\cal K}_N$. We start from an ansatz
\beq
n^{1-\text{loop,\;MHV}}_{\underline{1}|2|3|\ldots |N}(\ell) = n_{\underline{1}|2|3|\ldots|N}^0 + \ell_\mu n^\mu_{\underline{1}|2|3|\ldots|N} + \ell_\mu \ell_\nu n^{\mu \nu}_{\underline{1}|2|3|\ldots|N} + \ldots + \ell_{\mu_1} \ell_{\mu_2}  \ldots \ell_{\mu_{N-4}} n^{\mu_1\ldots \mu_{N-4}}_{\underline{1}|2|3|\ldots|N}
\label{loopnum18}
\eeq
incorporating the admissible powers ($\leq N-4$) of the loop momentum and convert this to a degree $N-4$ polynomial in proper times using the Schwinger parametrization $\ell^\mu \rightarrow L^\mu = \sum_{i=2}^N k_i^\mu \nu_i$\, ; see eq.~(\ref{loopnum3}). The latter must be compared with the worldline limits (\ref{loopnum10}) of the string integrands, i.e.~we impose
\beq
n^{1-\text{loop,\;MHV}}_{\underline{1}|2|3|\ldots |N}(\ell \rightarrow L) 
= K_N \, \Big|_{\te{MHV}}  \ ,
\label{loopnum19}
\eeq
where the MHV specification of $K_N$ translates its kinematic constituents $C_{1|A,B,C}$ to the expressions (\ref{loopnum17}). Each monomial in $\nu_j$ leads to a constraint on the symmetric tensors $n^{\mu_1\ldots \mu_{p}}_{\underline{1}|2|3|\ldots|N}$ with $0 \leq p \leq N-4$. Given that the four-dimensional vector indices range over $\mu_j=0,1,2,3$, the number of equations encoded in eq.~(\ref{loopnum19}) always suffices to determine all independent tensor components -- without any appearances of Gram determinants. In fact, a naive counting of equations and numerator degrees of freedom raises the possibility that the equation systems might be overconstrained. Hence, the existence of solutions which we verified for MHV helicities up to $N= 10$ can be viewed as a non-trivial consistency check on the underlying string integrands ${\cal K}_N$. That is why we conjecture the non-anomalous correlators of ref.~\cite{Mafra:2012kh} to completely capture the MHV sector of superstring one-loop amplitudes. For NMHV helicities, on the other hand, the BRST-invariant six-point correlator in eq.~(\ref{loopnum13}) turns out to be incompatible with numerators in eq.~(\ref{loopnum19}), calling for its anomalous completion.

At $N=4$, the four-point integrand in eq.~(\ref{loopnum7}) immediately determines the box numerator,
\beq
n^{0}_{\underline{1}|2|3|4}
= s_{23} s_{34} A^{\te{tree}}(1,2,3,4) \, \Big|_{\te{MHV}} 
 \ \ \ \Rightarrow \ \ \ \num_{\underline{1}|2|3|4} = X_{2,3} X_{3,4} \ ,
\label{loopnum20}
\eeq
so the simplest dependence on $\nu_j$ occurs at $N=5$ with ${\cal K}_5$ in eq.~(\ref{loopnum7}). In the decomposition (\ref{loopnum18}) of the pentagon numerator, its scalar part can be straightforwardly read off from
\begin{align}
n^{0}_{\underline{1}|2|3|4|5}& = K_5 \, \Big|^{\nu_j=0}_{\te{MHV}}  = \frac{s_{23}}{2} C_{1|23,4,5} + (23\leftrightarrow 24,25,34,35,45) \, \Big|_{\te{MHV}}   \ \
\Rightarrow \ \ \num_{\underline{1}|2|3|4|5} \, \Big|_{\ell^0}  \! &=  X_{2,3} X_{2,5} X_{4,5} \ ,
\label{loopnum21}
\end{align}
whereas the vector pentagon is determined by
\begin{align}
k^2_\mu n^{\mu}_{\underline{1}|2|3|4|5}& = K_5 \, \Big|^{\nu_2}_{\te{MHV}}  =  \big[ s_{23} \,C_{1|23,4,5} + (3\leftrightarrow 4,5) \big] \, \Big|_{\te{MHV}}
\label{loopnum22}
\end{align}
and its three independent images under $(2\leftrightarrow 3,4,5)$. As a unique solution, we find
\beq
\num_{\underline{1}|2|3|4|5}  \, \Big|_{\ell^1} 
 = X_{\ell,2} X_{3,5} X_{4,5}+X_{\ell,5} X_{2,3} X_{2,4} \ .
\label{loopnum23}
\eeq
Note that the MHV vector pentagon (\ref{loopnum23}) descends from a vectorial BRST invariant in pure-spinor superspace \cite{Mafra:2014oia, Mafra:2014gsa} which was identified as the vector pentagon in the five-point amplitude of ten-dimensional SYM \cite{Mafra:2014gja}.

Assembling eqs.~(\ref{loopnum21}) and (\ref{loopnum23}) reproduces the pentagon numerator in eq.~(\ref{5ptnu}). Note that the scalar part (\ref{loopnum21}) stems from $\partial G_{ij} \rightarrow \tfrac{1}{2} \te{sgn}_{ji}$ and depends on the ordering of the pentagon legs, e.g.~the numerator associated with ordering $\{1,2,4,3,5\}$ is obtained from eq.~(\ref{loopnum22}) by flipping the sign of $s_{34}C_{1|34,2,5}$. The vector pentagon (\ref{loopnum23}) due to $\partial G_{ij} \rightarrow \nu_{ij}$, on the other hand, is permutation invariant, i.e.~universal to all pentagon orderings. The analogous derivation of the hexagon numerator (\ref{hexnum}) using ${\cal K}_6$ in eq.~(\ref{loopnum13}) involves relations such as
\beq
k_{2\mu} k_{2\nu} n^{\mu \nu}_{\underline{1}|2|3|4|5|6}
=
- \big[ s_{23} \, s_{24} \, C_{1|324,5,6}  + (34\leftrightarrow 35,36,45,46,56) \big] \, \Big|_{\te{MHV}} \ .
\label{hextens}
\eeq
At the practical level, it proves convenient to convert the tensor numerators in the ansatz (\ref{loopnum18}) to spinorial expressions: Given that any $\ell$-dependence can be expressed in terms of structure constants $X_{\ell,\ldots}$, the left-handed spinor indices are entirely carried by $\lambda_1^\alpha\equiv \langle 1|$,
\beq
\num^{\mu_1\ldots \mu_{p}}_{\underline{1}|2|3|\ldots|N} = \sigma^{\mu_1}_{\alpha_1 \dot \beta_1} \sigma^{\mu_2}_{\alpha_2 \dot \beta_2} \ldots \sigma^{\mu_p}_{\alpha_p \dot \beta_p}  \num^{\alpha_1\ldots \alpha_{p}| \dot \beta_1 \ldots \dot \beta_p}_{\underline{1}|2|3|\ldots|N}  \co 
\num^{\alpha_1\ldots \alpha_{p}| \dot \beta_1 \ldots \dot \beta_p}_{\underline{1}|2|3|\ldots|N} = \lambda_1^{\alpha_1}  \lambda_1^{\alpha_2}  \ldots \lambda_1^{\alpha_p}  \widehat{\num}^{\dot \beta_1 \ldots \dot \beta_p}_{\underline{1}|2|3|\ldots|N} \ ,
\label{loopnum24}
\eeq 
where the tensors $\num^{\mu_1\ldots \mu_{p}}_{\underline{1}|2|3|\ldots|N}$ are defined by analogy with eq.~(\ref{loopnum18}). Hence, the leftover task to determine the right-handed part $\widehat{\num}^{\dot \beta_1 \ldots \dot \beta_p}_{\underline{1}|2|3|\ldots|N}$ involves fewer equations and unknowns.

\subsection{Reducible pieces}
\label{sec:poles}

The above procedure to determine the irreducible part of $N$-point one-loop amplitudes will now be complemented by a prescription to adjoin the kinematic poles $s_{i_1i_2\ldots i_p}^{-1}$ from external tree-level subdiagrams; see eq.~(\ref{mand}) for our conventions for Mandelstam invariants. The string-theory origin of the propagators (\ref{treenum3}) in tree-level amplitudes can be traced back to the representation
\beq
\delta(z) = \lim_{s\rightarrow 0} s z^{s-1} \ \ \ \leftrightarrow \ \ \ \int_0^a \dd z \, f(z) \, z^{s-1} = \frac{ f(0) }{s} + {\cal O}(s^0) \co a>0
\label{loopnum25}
\eeq
of the delta function. Similarly, one-loop amplitudes as in eq.~(\ref{loopnum6}) develop kinematic poles from the local behaviour of $\partial {\cal G}(z_i,z_j,\tau) e^{-2{\cal Q}_N} \sim |z_i - z_j|^{\alpha' s_{ij} - 1}$ as $z_i \rightarrow z_j$. However, the factor of $|z_i - z_j|^{\alpha' s_{ij} - 1}$ can only integrate to a pole in $s_{ij}$ if $i$ and $j$ are neighbors in the cyclic integration domain $\Im(z_i) < \Im(z_{i+1})$ of the open string. These observations can be summarized by the following behaviour in the field-theory limit:
\beq
\pa {\cal G}(z_i,z_j,\tau)  \rightarrow \pm \frac{\delta_{i,j\mp 1} \, \de(z_{i}-z_{j}) }{s_{ij}}  + \partial G_{ij} \ .
\label{loopnum26}
\eeq
The prescription (\ref{loopnum26}) yields the following reducible contributions in the five-point integrand (\ref{loopnum7}):
\beq
{\cal K}_5 \, \Big|_{\te{reducible}} =  \delta(z_2-z_3)  C_{1|23,4,5} +  \delta(z_3-z_4)  C_{1|34,2,5} +  \delta(z_4-z_5)  C_{1|45,2,3} \ .  
\label{loopnum27}
\eeq
As known from refs.~\cite{Bern:1990cu,Bern:1991aq},
it is crucial to identify these reducible parts prior to the degeneration of the Green functions in eq.~(\ref{loop3d}) as the worldline Green function $\partial G_{ij}$ does not preserve the singularity of the worldsheet Green function $\partial {\cal G}(z_i,z_j,\tau)\sim(z_i-z_j)^{-1}$. In the worldline limit of the reducible part (\ref{loopnum27}), the delta functions are translated via $\delta(z_i-z_j)= t^{-1} \delta(\nu_i-\nu_j)$. Hence, both the overall power of $t$ and the momentum-dependence of $Q_5 |_{\nu_i=\nu_j}$ agree with a box integral (\ref{loopnum1}) with momentum $k_i+k_j$ in a massive corner. The reducible part of the correlator in eq.~(\ref{loopnum27}) identifies $s_{45} C_{1|45,2,3}$ to be the corresponding box numerator\footnote{The factor of $s_{45}$ can be understood from an insertion of $1=\frac{s_{45}}{s_{45}}$ into the contribution $\sim \frac{ C_{1|45,2,3} }{\ell^2 (\ell+k_1)^2 (\ell+k_{12})^2 (\ell + k_{123})^2}$ of this box diagram to the integrand of the one-loop SYM amplitude.}, in agreement with eq.~(\ref{5ptnu}).

Note that by construction of the string integrands ${\cal K}_N$, integration by parts identities such as eq.~(\ref{loopnum11}) have been used to eliminate any instance of $\pa {\cal G}(z_1,z_j,\tau)$ with $j=2,3,\ldots,N$. Hence, $z_1$ will never appear in a delta function in eq.~(\ref{loopnum27}), and the field-theory limit cannot comprise any Feynman integrals with leg 1 in a massive corner. This is the string-theory origin of the selection rule on the BCJ numerators which was earlier on motivated by the vanishing of the associated kinematic structure constant, $X_{1,j}=0$.

\begin{figure}
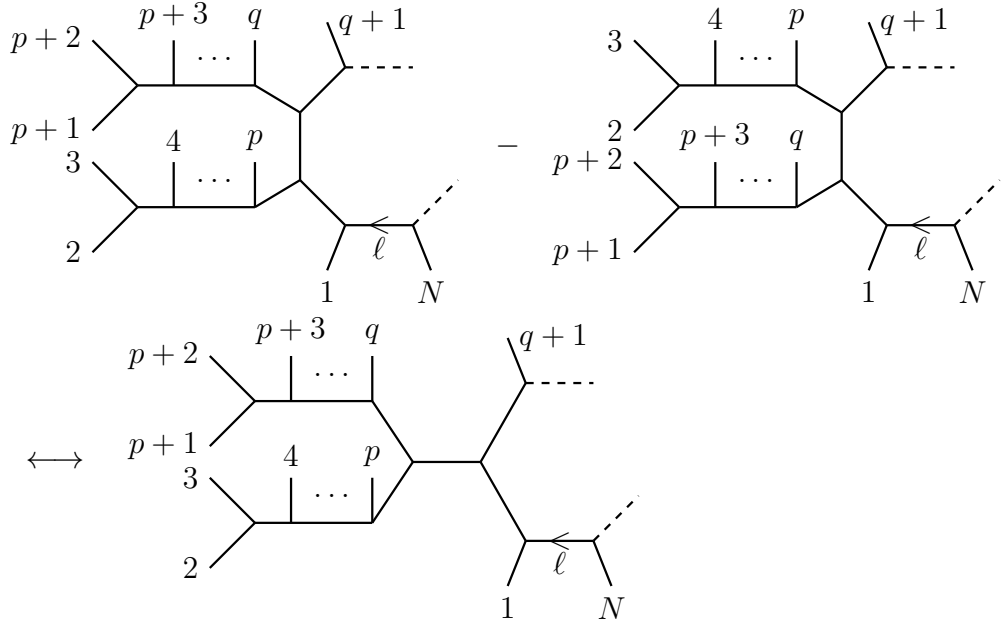

\begin{center}
\tikzpicture [scale=1.2,line width=0.30mm]
\draw (1.8,0.875) node{$-$};
\draw (-3.1,-2.625) node{$\longleftrightarrow \ \ $};
\scope[xshift=2cm,yshift=-3.5cm]
\draw (0,0) -- (0.75,0) ;
\draw (0,0) -- (-0.5,0.875) ;
\draw (-0.5,0.875) -- (-1.25,0.875);
\draw (-0.5,0.875) -- (0,1.75);
\draw[dashed] (0,1.75) -- (0.75,1.75) ;
\draw[dashed] (0.75,0) -- (1.25,0.5) ;
\draw (0.375,0) node{$<$} node[below]{$\ell$};
\draw (0,0) -- (-0.2,-0.5)node[below]{$1$} ;
\draw (-1.25,0.875) -- (-1.7,1.55);
\draw (-1.7,1.55)--(-3,1.55);
\draw (-3,1.55) -- (-3.5,2.05) node[left]{$p+2$};
\draw (-3,1.55) -- (-3.5,1.05) node[left]{$p+1$};
\draw (-2.15,1.85) node{$\ldots$};
\draw (-2.6,1.55) -- (-2.6,2.05) node[above]{$p+3$};
\draw (-1.7,1.55) -- (-1.7,2.05) node[above]{$q$};
\draw (-1.25,0.875) -- (-1.7,0.2);
\draw (-1.7,0.2)  -- (-3,0.2);
\draw (-2.6,0.2) -- (-2.6,0.7) node[above]{4};
\draw (-2.15,0.5) node{$\ldots$};
\draw (-3,0.2) -- (-3.5,0.7) node[left]{3};
\draw (-3,0.2)  -- (-3.5,-0.3) node[left]{2};
\draw (-1.7,0.2)--(-1.7,0.7) node[above]{$p$};
\draw (0,1.75) -- (-0.2,2.25) node[right]{$q+1$};
\draw (0.75,0) -- (0.95,-0.5) node[below]{$N$};
\endscope
\begin{scope}[xshift= 0cm,yshift=0cm]
\draw (0,0) -- (0.75,0) ;
\draw (0,0) -- (-0.5,0.5) ;
\draw (-0.5,0.5) -- (-0.5,1.25);
\draw (-0.5,1.25) -- (0,1.75);
\draw[dashed] (0,1.75) -- (0.75,1.75) ;
\draw[dashed] (0.75,0) -- (1.25,0.5) ;
\draw (0.375,0) node{$<$} node[below]{$\ell$};
\draw (0,0) -- (-0.2,-0.5)node[below]{$1$} ;
\draw (-0.5,0.5) -- (-1,0.2) ;
\begin{scope}[xshift=0.7cm]
\draw (-1.7,0.2)  -- (-3,0.2);
\draw (-2.6,0.2) -- (-2.6,0.7) node[above]{4};
\draw (-2.15,0.5) node{$\ldots$};
\draw (-3,0.2) -- (-3.5,0.7) node[left]{$3$};
\draw (-3,0.2)  -- (-3.5,-0.3) node[left]{$2$};
\draw (-1.7,0.2)--(-1.7,0.7) node[above]{$p$};
\end{scope}
\draw (-0.5,1.25) -- (-1,1.55);
\begin{scope}[xshift=0.7cm]
\draw (-1.7,1.55)--(-3,1.55);
\draw (-3,1.55) -- (-3.5,2.05) node[left]{$p+2$};
\draw (-3,1.55) -- (-3.5,1.05) node[left]{$p+1$};
\draw (-2.15,1.85) node{$\ldots$};
\draw (-2.6,1.55) -- (-2.6,2.05) node[above]{$p+3$};
\draw (-1.7,1.55) -- (-1.7,2.05) node[above]{$q$};
\end{scope}
\draw (0,1.75) -- (-0.2,2.25) node[right]{$q+1$};
\draw (0.75,0) -- (0.95,-0.5) node[below]{$N$};
\end{scope}
\begin{scope}[xshift= 6cm]
\draw (0,0) -- (0.75,0) ;
\draw (0,0) -- (-0.5,0.5) ;
\draw (-0.5,0.5) -- (-0.5,1.25);
\draw (-0.5,1.25) -- (0,1.75);
\draw[dashed] (0,1.75) -- (0.75,1.75) ;
\draw[dashed] (0.75,0) -- (1.25,0.5) ;
\draw (0.375,0) node{$<$} node[below]{$\ell$};
\draw (0,0) -- (-0.2,-0.5)node[below]{$1$} ;
\draw (-0.5,0.5) -- (-1,0.2);
\begin{scope}[xshift=0.7cm,yshift=1.35cm]
\draw (-1.7,0.2)  -- (-3,0.2);
\draw (-2.6,0.2) -- (-2.6,0.7) node[above]{4};
\draw (-2.15,0.5) node{$\ldots$};
\draw (-3,0.2) -- (-3.5,0.7) node[left]{$3$};
\draw (-3,0.2)  -- (-3.5,-0.3) node[left]{$2$};
\draw (-1.7,0.2)--(-1.7,0.7) node[above]{$p$};
\end{scope}
\draw (-0.5,1.25) -- (-1,1.55);
\begin{scope}[xshift=0.7cm,yshift=-1.35cm]
\draw (-1.7,1.55)--(-3,1.55);
\draw (-3,1.55) -- (-3.5,2.05) node[left]{$p+2$};
\draw (-3,1.55) -- (-3.5,1.05) node[left]{$p+1$};
\draw (-2.15,1.85) node{$\ldots$};
\draw (-2.6,1.55) -- (-2.6,2.05) node[above]{$p+3$};
\draw (-1.7,1.55) -- (-1.7,2.05) node[above]{$q$};
\end{scope}
\draw (0,1.75) -- (-0.2,2.25) node[right]{$q+1$};
\draw (0.75,0) -- (0.95,-0.5) node[below]{$N$};
\end{scope}
\endtikzpicture
\end{center}
\caption{Given the impact of worldsheet Green functions on the field-theory limit, antisymmetrized $p$-gon numerators agree with $(p-1)$-gon numerators as long as leg 1 adjacent to the loop momentum is not involved in the antisymmetrization.}
\label{NNminus1}
\end{figure}

We would like to highlight the two-fold role of the worldsheet Green function $\pa {\cal G}(z_i,z_{i+1},\tau)$. On the one hand, by its contribution $\te{sgn}_{i,i+1}$ from the worldline limit (\ref{loop3d}), it controls the antisymmetric part of the $N$-gon numerator under exchange of legs $i,i+1$. On the other hand, its reducible contribution (\ref{loopnum26}) implies that the antisymmetric part of the $N$-gon numerator coincides with the corresponding $(N-1)$-gon numerator where legs $i$ and $i+1$ form an external tree. Generally speaking, this two-fold role of the worldsheet Green function offers a string-theory understanding of kinematic Jacobi relations between $m$-gons and $(m-1)$-gons as depicted in figure \ref{NNminus1}. However, a separate analysis is required if leg 1 adjacent to the loop momentum is involved in this antisymmetrization, see the discussion around eq.~(\ref{hardjac}).

\subsection{Multiparticle pattern of reducible pieces}

In presence of $N \geq 6$ external legs, multiple factors of worldsheet Green functions allow for simultaneous kinematic poles such as $(s_{23}s_{45})^{-1}$. Their overall number $p$ can vary in the range $0\leq p \leq N-4$. Iteration of the five-point prescription (\ref{loopnum26}) gives rise to the following pole structure in the field-theory limit:
\begin{align}
&\pa {\cal G}(z_2,z_3,\tau) \pa {\cal G}(z_4,z_5,\tau)  \rightarrow \frac{\delta(z_2-z_3) \delta(z_4-z_5)}{s_{23} s_{45}} \label{loopnum28} \\
& \ \ \ +\frac{\delta(z_2-z_3) \pa G_{45}}{s_{23} } 
+\frac{ \pa G_{23} \delta(z_4-z_5)}{s_{45}} 
+\pa G_{23} \pa G_{45} \ .
\notag
\end{align}
Nested arguments of the Green functions lead to multiparticle Mandelstam invariants, e.g.
\begin{align}
\pa {\cal G}(z_2,z_3,\tau)  \pa {\cal G}(z_2,z_4,\tau) 
\rightarrow  \frac{\delta(z_2-z_3) \delta(z_3-z_4) }{s_{234} s_{23}} + \frac{\delta(z_2-z_3) \pa G_{24}  }{s_{23}}
 +\pa G_{23} \pa G_{24} \ . \label{loopnum29}
\end{align}
If only a subset of the Green functions refer to a sequence of neighboring particles in the integration domain $z_i \leq z_{i+1}$, then the cascade of reducible contributions gets shortened, e.g.
\begin{align}
\pa {\cal G}(z_2,z_3,\tau)  \pa {\cal G}(z_3,z_5,\tau) 
\rightarrow   \frac{\delta(z_2-z_3)  \pa G_{35}  }{s_{23}}
 +\pa G_{23}   \pa G_{35}  \ . \label{loopnum30}
\end{align}
The all-multiplicity generalization of the above rules closely follows the pole analysis at tree level; see in particular section 4 of ref.~\cite{Broedel:2013tta}.

The following types of six-point numerators arise from the integrand in eq.~(\ref{loopnum13}) as well as the prescriptions (\ref{loopnum28}) to (\ref{loopnum30}):
\begin{itemize}
\item massive pentagons from ${\cal G}(z_2,z_3,\tau) \rightarrow \delta(z_2-z_3)/ s_{23}$ along with
\begin{align}
& \! \! \! \! \! \! \! \!  \! \! \! \! \! \! \!  {\cal K}_6 \, \Big|_{\partial {\cal G}(z_2,z_3,\tau)} = \big[ Y_{34} C_{1|234,5,6}- Y_{24} C_{1|324,5,6} + (4\leftrightarrow 5,6) \big] + \big[ Y_{45} C_{1|23,45,6} + (45\leftrightarrow 46,56)\big] 
\label{loopnum31}
\end{align}
The numerator is determined by applying the procedure in subsection \ref{sec:irred} to eq.~(\ref{loopnum31}) with $z_2=z_3$ or $\nu_2=\nu_3$, resulting in the following contribution to the color-ordered SYM integrand:
\beq
\frac{ \delta^8(Q)}{\prod_{j=2}^6 \langle 1j\rangle^2} \frac{ X_{2+3,5}X_{2+3,4}X_{\ell,6}+X_{2+3,6}X_{2+3,4}X_{2+3+4,5}+X_{4,6}X_{\ell,2+3}X_{2+3+4,5} }{ \ell^2 (\ell+k_1)^2 (\ell + k_{123})^2 (\ell + k_{1234})^2 (\ell + k_{12345})^2} \frac{X_{2,3} }{s_{23}} \ .
\eeq
\item two-mass boxes from $Y_{23}Y_{45}C_{1|23,45,6} ,Y_{23}Y_{56}C_{1|23,4,56} $ and $Y_{34}Y_{56}C_{1|2,34,56} $, e.g.
\begin{align}
&\!\!\!\!\!\!\! \frac{ C_{1|23,45,6} }{\ell^2 (\ell+k_1)^2 (\ell + k_{123})^2 (\ell + k_{12345})^2} \, \Big|_{\te{MHV}} = \frac{ \delta^8(Q)}{\prod_{j=2}^6 \langle 1j\rangle^2}
 \frac{  X_{2+3,4+5} X_{4+5,6} }{ \ell^2 (\ell+k_1)^2 (\ell + k_{123})^2 (\ell + k_{12345})^2}  \frac{X_{2,3} X_{4,5} }{s_{23} s_{45}}
 \label{massb1}
\end{align}
\item one-mass boxes from $Y_{23}(Y_{24}+Y_{34}) C_{1|234,5,6} + (3\leftrightarrow 4)$ and $(234\leftrightarrow 345,456)$, e.g.
\begin{align}
& \frac{ C_{1|234,5,6} }{\ell^2 (\ell+k_1)^2 (\ell + k_{1234})^2 (\ell + k_{12345})^2} \, \Big|_{\te{MHV}} \notag \\
& \ \ \ \ \ = \frac{ \delta^8(Q)}{\prod_{j=2}^6 \langle 1j\rangle^2}
\frac{   X_{2+3+4,5} X_{5,6} }{ \ell^2 (\ell+k_1)^2 (\ell + k_{1234})^2 (\ell + k_{12345})^2} \, \left( 
 \frac{ X_{2,3} X_{2+3,4} }{s_{23} s_{234}} +   \frac{ X_{4,3} X_{4+3,2}  }{s_{34} s_{234}}
\right)
 \label{massb2}
\end{align}
\end{itemize}
The complete color-ordered six-point integrand comprising various permutations of the above diagrams is spelt out in appendix \ref{app:examples}. Note that the above box and pentagon numerators descend from the ten-dimensional one-loop SYM amplitude in ref.~\cite{Mafra:2014gja} upon dimensional reduction to $D=4$ and specialization to MHV helicities.

The rewriting of the box contributions in eqs.~(\ref{massb1}) and (\ref{massb2}) exemplifies the conversion between cubic diagrams and Parke-Taylor-like denominators in eq.~(\ref{loopnum17}) for $C_{1|A,B,C}$, \eqref{massb3} and \eqref{massb4}. This ensures that the box contribution of the $N$-point one-loop integrand,
\beq
 \sum_{2\leq p < q \leq N-1} \frac{ C_{1|23\ldots p , p+1\ldots q , q+1\ldots N} }{\ell^2 (\ell+k_1)^2 (\ell+k_{12\ldots p})^2 (\ell+k_{12\ldots p,p+1\ldots q})^2} \ ,
\label{massb5}
\eeq
comprises all the tree-level subdiagrams admissible in the massive corners of the boxes. The sum over diagrams in eq.~(\ref{massb4}) ties in with the construction of $C_{1|A,B,C}$ in pure-spinor superspace, where supersymmetric Berends-Giele currents enter as a key ingredient \cite{Mafra:2012kh, Mafra:2014oia}. Hence, eq.~(\ref{massb5}) is expected to capture the box content of ten-dimensional $N$-point SYM amplitudes in pure-spinor superspace, in line with the five- and six-point results in ref.~\cite{Mafra:2014gja}.

\section{Conclusion and outlook}\label{sec6}

In this work, we present remarkably compact expressions for BCJ numerators of one-loop MHV amplitudes in maximally supersymmetric Yang-Mills theory and supergravity. While the representation for supergravity amplitudes is obtained via the BCJ double-copy construction~\cite{Bern:2008qj,Bern:2010ue}, that for SYM is derived from the infinite-tension limit of open superstrings. The main results for SYM numerators are given in eq.~(\ref{massgen}) and can be produced by a simple operation $\mathscr{X}$  acting on the self-dual numerators built from kinematic structure constants. 

The underlying superstring correlators have been partially determined in ref.~\cite{Mafra:2012kh} by imposing BRST invariance of the superstring. Still, the fingerprints of the hexagon gauge anomaly remain to be incorporated, presumably by combining the kinematic factors constructed in ref.~\cite{Mafra:2014gsa} with additional worldsheet functions $f^{(n)}$ with $n=2,3,\ldots$ described in ref.~\cite{Broedel:2014vla}. Fortunately, when restricted to the MHV helicity configuration in four dimensions, we find strong evidence that the non-anomalous correlators in ref.~\cite{Mafra:2012kh} appear to completely capture the BCJ numerators of this work. 

The natural next step is to extend our one-loop numerators to arbitrary helicity configurations and to higher dimensions. Five- and six-point generalizations are already available in ten-dimensional pure-spinor superspace~\cite{Mafra:2014gja}. It would be interesting to extend these superspace numerators to arbitrary multiplicity and to incorporate the BCJ duality into this framework beyond five points. Also, it would be desirable to understand the four-dimensional helicity selection rules associated with the ten-dimensional hexagon anomaly, i.e.~to make the vanishing of anomalous kinematic factors \cite{Mafra:2014gsa} in the MHV helicity sector\footnote{We would like to thank Carlos Mafra for contributing to these checks.} more transparent.

Although the explicit form of the numerators is worked out here only for the MHV case, we emphasize that the diagrammatic structure of our representation as derived from string theory is dimension-agnostic: No integral reductions have been performed to eliminate certain $m$-gon integrals or loop momenta in the numerators, and the structure extends to arbitrary helicities and higher dimensions. Thus our representation of SYM amplitudes differs significantly from other forms in four dimensions, e.g.~those obtained using generalized unitarity~\cite{Bern:1994zx,Bern:1994cg,Britto:2004nc}, though it should be possible to show their equivalence by reductions of higher-gon integrals~\cite{Bern:1992em}. It also takes a different form from previous BCJ representations at five \cite{Carrasco:2011mn} and six points~\cite{Bjerrum-Bohr:2013iza}; most notably, the only kinematic poles of our numerators are introduced by the polarization vectors.

The virtue of these BCJ representations are that one gets supergravity integrands for free, in particular at one loop it takes the very enlightening form in eq.~(\ref{sugra}) upon combining the master numerators (\ref{gen_result}) with any representation of the color-ordered SYM integrand. As discussed in ref.~\cite{Bern:2011rj}, given Yang-Mills integrands with $0\leq {\cal N}\leq 4$ supersymmetries, our result (\ref{gen_result}) immediately yields a new, explicit representation for supergravity amplitudes with ${\cal N}+4$ supersymmetries.

The simplicity of our result and its connections to the self-dual sector strongly suggest that it is possible to assemble MHV numerators at higher loops from the same building blocks. The superstring could again be a valuable starting point, together with the techniques of ref.~\cite{Tourkine:2013rda} to recover Feynman integrals. Pure-spinor superspace techniques provide BCJ-satisfying two-loop five-point numerators \cite{Mafra:2015mja}, and the choice of building blocks is inspired by the low-energy analysis of the underlying superstring amplitude \cite{Gomez:2015uha}. In an extension to higher multiplicity, one expects significant simplifications in the MHV sector, and it would be very interesting to find an analogue of the operation $\mathscr{X}$, which maps self-dual numerators to MHV ones, at higher loops. A two-loop relation  found in ref.~\cite{Badger:2013gxa}, similar to the dimension-shifting formula, provides an important clue.

To conclude, we point out an interesting connection. Self-dual gauge theory is known to be classically integrable, and this property has been one of the main motivations for its study. We find in this work an intriguing connection between this theory and ${\cal N}= 4$ SYM, a theory which is thought to be quantum integrable in the planar limit (at one loop, the planar part determines the complete amplitude). Perhaps it is not surprising that we find the same kinematic structures in both theories. It would be important to understand the full extent of this connection.

\subsection*{Acknowledgments}

We are very grateful to Nima Arkani-Hamed, Zvi Bern, Henrik Johansson, Carlos Mafra, Donal O'Connell, Stephan Stieberger, Piotr Tourkine and Pierre Vanhove for valuable discussions. In addition, we would like to thank Carlos Mafra and Donal O'Connell for collaboration on related topics as well as Piotr Tourkine for insightful comments on the draft. Research at Perimeter Institute is supported by the Government of
Canada through Industry Canada and by the Province of Ontario through the Ministry of
Research \& Innovation. RM is a Marie Curie fellow and a JRF at Linacre College. OS acknowledges financial support by the European Research Council Advanced Grant No.~247252 of Michael Green. SH and OS would like to thank Simons Center for Geometry and Physics and the organizers of the workshop ``Mathematics and Physics of Scattering Amplitudes" for their hospitality, where a preliminary version of this work was done. RM and SH are grateful to the Albert-Einstein-Institute for hospitality during advanced stages of this work.

\appendix

\section{String-theory derivation of tree-level MHV numerators}\label{app:string_tree}

In this appendix, we derive the tree-level MHV numerators from the field-theory limit ($\alpha' \rightarrow 0$) of string theory. Supersymmetric gauge theories arise from the massless excitations of open superstrings, and their tree-amplitudes are obtained by integrating over worldsheets of disk topology. 

\subsection{The ten-dimensional superstring origin}
We recall the representation of the integrand given in ref.~\cite{Mafra:2011kj} to extract BCJ numerators in ten-dimensional pure-spinor superspace. Up to total derivatives in the integration variables $z_2,\ldots,z_{N-2}$, this is the genus-zero worldsheet correlator $\langle \! \langle\ldots \rangle \! \rangle$ of $N$ (un-)integrated vertex operators $V_i,U_i$ of the gauge supermultiplet in the pure-spinor formalism \cite{Berkovits:2000fe},
\begin{align}
&\langle \! \langle V_1(z_1) U_2(z_2)\cdots U_{N-2}(z_{N-2}) V_{N-1}(z_{N-1})V_N(z_N) \rangle \! \rangle \sim \prod_{i<j} |z_{i}-z_{j}|^{\alpha' k_{i}\cdot k_{j}}  \label{treenum1}
\\
&\ \ \times 
\sum_{j=1}^{N-2} \frac{\langle V_{12\ldots j} V_{N-1,N-2\ldots j+1} V_N \rangle}{(1,2,3,\ldots,j,N,j+1,\ldots,N-2,N-1)} + \textrm{perm}(2,3,\ldots,N-2) \ .\notag
\end{align}
The kinematic building blocks $V_{12\ldots p}$ encompassing several particles $1,2,\ldots,p$ are known superfields incorporating gluon polarizations, gluino wave functions and momenta. Representations for $V_{12\ldots p}$ were originally obtained from an iterated operator product expansion of the vertex operators in eq.~(\ref{treenum1})\footnote{Note that the precursors of $V_{12\ldots p}$ were denoted by $T_{12\ldots p}$ in these references.} \cite{Mafra:2010ir, Mafra:2010jq, Mafra:2010gj, Mafra:2011kj}. Later on, they were embedded into a more general and computationally efficient formalism of multiparticle superfields \cite{Mafra:2014oia} which build up solutions of non-linear field equations of ten-dimensional SYM \cite{Mafra:2015gia}. Using the multiparticle version $V_{12\ldots p}$ of the unintegrated vertex operator \cite{Mafra:2014oia}, the final result in eq.~(\ref{treenum1}) for the correlator can be rewritten in terms of $(N-2)!$ permutations of 
\beq
(1,2,\ldots,k-1,k) \equiv (z_1-z_2)(z_2-z_3) \cdots (z_{k-1}-z_k)(z_k-z_1)\ ,
\label{treenum2}
\eeq
after undoing the SL$_2$ fixing of ref.~\cite{Mafra:2011kj}. The bracket $\langle \ldots\rangle$ in the second line of eq.~(\ref{treenum1}) instructs us to extract the fifth power of the Grassmann variable $\theta$ from the enclosed superfields \cite{Berkovits:2000fe}; see ref.~\cite{Mafra:2010pn} for a computer implementation and ref.~\cite{PSwebsite} for various bosonic and fermionic components of the resulting SYM amplitudes \cite{Mafra:2010jq}.

\begin{figure}[h]
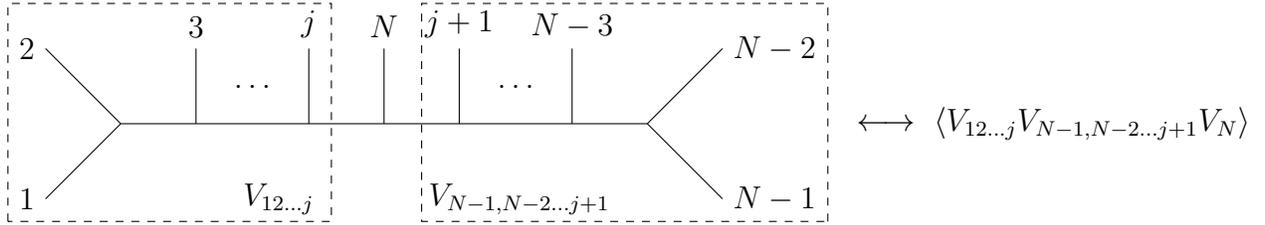

\begin{center}
\tikzpicture
\draw(0,0) --(-1,-1) node[left]{$1$};
\draw(0,0) --(-1,1) node[left]{$2$};
\draw(0,0) -- (7,0);
\draw(1,0) -- (1,1) node[above]{$3$};
\draw (1.75,0.5)node{$\ldots$};
\draw (2.5,0)--(2.5,1) node[above]{$j$};
\draw (3.5,0)--(3.5,1) node[above]{$N$};
\draw (4.5,0)--(4.5,1) node[above]{$j+1$};
\draw (5.25,0.5)node{$\ldots$};
\draw(6,0) -- (6,1) node[above]{$N-3$};
\draw(7,0) --(8,-1) node[right]{$N-1$};
\draw(7,0) --(8,1) node[right]{$N-2$};
\draw[dashed] (-1.5,-1.3) rectangle (2.8,1.6);
\draw[dashed] (9.4,-1.3) rectangle (4.0,1.6);
\draw(2.1,-1)node{$V_{12\ldots j}$};
\draw(5.3,-1)node{$V_{N-1,N-2\ldots j+1}$};
\draw(12.4,0)node{$\longleftrightarrow  \ \langle V_{12\ldots j} V_{N-1,N-2\ldots j+1} V_N \rangle $};
\endtikzpicture
\end{center}
\caption{The mapping between master numerators and expressions in pure-spinor superspace.}
\label{PSS}
\end{figure}

The kinematic factors $\langle V_{12\ldots j} V_{N-1,N-2\ldots j+1} V_N \rangle$ furnish master numerators under kinematic Jacobi relations \cite{Bern:2008qj} associated with $(N-2)!$ cubic half-ladder diagrams, see figure \ref{PSS}. Their endpoints $1$ and $N-1$ are inherited from the unintegrated vertices at finite positions (say $z_1=0$ and $z_{N-1}=1$), whereas the role of $V_N$ is inherited from the SL$_2$ fixing $z_N\rightarrow \infty$. The propagators of the cubic diagrams emerge in the $\alpha' \rightarrow 0$ limit of the following disk integrals\footnote{The SL$_2$ redundancy of a disk worldsheet can be removed by fixing three positions as $(z_i,z_j,z_k) \rightarrow (0,1,\infty)$ and adjoining the Jacobian factor $(z_{i}-z_j)(z_{i}-z_k)(z_{j}-z_k)$.} \cite{Mafra:2011nw, Broedel:2013tta},
\beq
\lim_{\alpha' \rightarrow 0} \int \limits_{z_{\sigma(i)}\leq z_{\sigma(i+1)}} \frac{ \dd z_1\, \dd z_2 \, \ldots \, \dd z_N}{\te{vol(SL$_2$)}} \frac{ \prod_{i<j} |z_{i}-z_{j}|^{\alpha' k_{i}\cdot k_{j}} }{(\rho(1),\rho(2),\ldots,\rho(N)) } = m[\sigma|\rho] \co \sigma,\rho \in S_N \ .
\label{treenum3}
\eeq
The doubly partial amplitude $m[\sigma|\rho]$ \cite{Cachazo:2013iea} comprises the $N-3$ simultaneous propagators of all the cubic diagrams compatible with the cyclic orderings $\sigma$ and $\rho$ set by the integration domain and the integrand of the form (\ref{treenum2}), respectively, for instance
\begin{align}
m[1,2,3,4\, | \, 1,2,3,4]&= \frac{1}{s_{12}}+\frac{1}{s_{23}} \co m[1,2,3,4\, | \, 1,2,4,3]=- \frac{1}{s_{12}} \\
m[1,2,3,4,5\, | \, 2,1,4,3,5]&= \frac{1}{s_{12}s_{34}} \co \! \! \! m[1,2,3,4,5\, | \, 1,3,5,2,4] = 0 \ .
\end{align}
By combining the correlation function (\ref{treenum1}) with the integration prescription (\ref{treenum3}), we arrive at the following cubic-graph organization of color-ordered tree amplitudes
\begin{align}
&A^{\te{tree}}(1,2,\ldots,N) = \lim_{\alpha' \rightarrow 0} \! \int \limits_{z_{i}\leq z_{i+1}} \! \frac{ \dd z_1\, \dd z_2 \, \ldots \, \dd z_N}{\te{vol(SL$_2$)}}  \langle \! \langle V_1(z_1) U_2(z_2)\cdots U_{N-2}(z_{N-2}) V_{N-1}(z_{N-1})V_N(z_N) \rangle \! \rangle
\notag \\
& = \sum_{j=1}^{N-2} \sum_{\rho \in S_{N-3}}   \langle V_{1\rho(2)\ldots \rho(j)} V_{N-1,\rho(N-2)\ldots \rho(j+1)} V_N \rangle  \label{treenum4} \\
& \ \ \ \ \ \ \ \times m[1,2,\ldots,N|1,\rho(2),\ldots,\rho(j),N,\rho(j+1),\ldots,\rho(N-2),N-1]  \ . \notag
\end{align}
Once the entire Kleiss-Kuijf basis $\{A^{\te{tree}}(1,\sigma(2),\ldots,\sigma(N-1),N) ,\sigma\in S_{N-2}\}$ of partial amplitudes has been evaluated via eq.~(\ref{treenum4}) with $m[1,2,\ldots,N-1,N|\ldots]\rightarrow m[1,\sigma(2),\ldots,\sigma(N-1),N|\ldots]$, the numerator for any cubic graph can be read off as a linear combination of the $(N-2)!$ master numerators $\langle V_{1\rho(2)\ldots \rho(j)} V_{N-1,\rho(N-2)\ldots \rho(j+1)} V_N \rangle$. These numerators solve all the tree-level kinematic Jacobi relations (\ref{jacob}); see ref.~\cite{Mafra:2011kj} for further details.

\subsection{Restriction to four-dimensional MHV kinematics}

The tree-level MHV numerators (\ref{treenumA}) in terms of kinematic structure constants can be reproduced from the expressions in eq.~(\ref{treenum4}) for ten-dimensional SYM amplitudes. The gluonic component of the local superspace numerators $\langle V_{12\ldots j} V_{N-1,N-2\ldots j+1} V_N \rangle$ can be extracted via ref.~\cite{Mafra:2010pn}\footnote{We are grateful to Carlos Mafra for providing the component expansions.}, and easily written in spinor-helicity variables for four-dimensional MHV helicities. 

With negative helicities in legs 1 and 2 and positive helicities in the remaining legs, it is straightforward to obtain the following master numerators:
\begin{align}
\langle V_{1} V_{2} V_{3} \rangle \, \MHV &= \frac{ \langle 1 2 \rangle^3 [\eta 2]}{[\eta 1] \langle 1 3 \rangle^2} = \frac{ \langle 1 2 \rangle^3}{\langle 2 3 \rangle\langle 3 1 \rangle}
\co &\langle V_{123} V_{4} V_{5} \rangle \, \MHV = \frac{ \langle 1 2 \rangle^4 [2\eta] [23] [45]}{[\eta 1]  \langle 13 \rangle \langle 14 \rangle \langle 15 \rangle} \,,
\\
\langle V_{12} V_{3} V_{4} \rangle \, \MHV &=  \frac{ \langle 1 2 \rangle^3 [2\eta] [34]}{[\eta 1]  \langle 1 3 \rangle\langle 1 4 \rangle } 
\co& \langle V_{12} V_{43} V_{5} \rangle \, \MHV = \frac{ \langle 1 2 \rangle^4 [2\eta] [25] [43]}{[\eta 1]  \langle 13 \rangle \langle 14 \rangle \langle 15 \rangle} \,,  \\
\langle V_{1} V_{32} V_{4} \rangle \, \MHV &= \frac{ \langle 1 2 \rangle^3 [\eta 4] [23]}{[\eta 1] \langle 1 3 \rangle\langle 1 4 \rangle }
\co& \langle V_{1} V_{432} V_{5} \rangle \, \MHV = \frac{ \langle 1 2 \rangle^3 [5 \eta] [25] [34]}{[\eta 1]\langle 13 \rangle \langle 14 \rangle}  \,.
\end{align}
These expressions are based on the polarization vectors in eq.~(\ref{treenum5}), reproduce the numerators in eq.~(\ref{treenumA}) and are conjectured to generalize to
\begin{align}
\langle V_{12\ldots j} V_{N-1,N-2,\ldots,j+1} V_{N} \rangle \, \MHV &= \left\{ \begin{array}{cl} \displaystyle \frac{(-1)^N \langle 12 \rangle^2 [N \eta ] \langle 1N \rangle }{[\eta 1] \prod_{i=3}^{N} \langle 1 i \rangle^2} \left( \prod_{q=1}^{N-3} X_{1+N+2+3+\ldots+q,q+1} \right)
 &: \ j=1  \ , \\
\displaystyle \frac{(-1)^N \langle 12 \rangle^3 [2\eta ] }{[\eta 1] \prod_{i=3}^{N} \langle 1 i \rangle^2} \left( \prod_{p=3}^{j} X_{1+2+\ldots+(p-1),p} \right) X_{1+2+\ldots+j,N}  & \\ \displaystyle
 \ \ \ \ \ \ \times \left( 
\prod_{q=1}^{N-j-2} X_{1+2+\ldots+j+N+(j+1)+\ldots+(j+q-1),j+q} \right) &: \ j\geq 2 \ ;
\end{array}\right.
\label{conjtree}
\end{align}
see figure \ref{PSS} for the underlying half-ladder diagram. We have explicitly checked eq.~(\ref{conjtree}) for six-point numerators, and there is no fundamental obstruction to extending the checks to higher multiplicity. 

Given that only $N-3$ legs enter the master numerators $\langle V_{1\rho(2)\ldots \rho(j)} V_{N-1,\rho(N-2)\ldots \rho(j+1)} V_N \rangle$ in a permutation-agnostic manner, the BCJ numerators determined by eq.~(\ref{treenum4}) violate crossing symmetry, i.e.~they treat legs $1,N-1,N$ associated with unintegrated vertex operators on special footing. This amounts to a particular distribution of contact terms -- the fingerprints of the quartic Feynman vertex -- among the cubic diagrams. As a common feature of contact terms, they involve at least two factors of $(\epsilon_i \cdot \epsilon_j)$ in terms of $D$-dimensional polarization vectors. On the other hand, tensor structures of the schematic form $(\epsilon_i \cdot \epsilon_j) (\epsilon_p \cdot k_q)^{N-2}$ are unambiguously tied to a specific cubic diagram and therefore furnish a crossing-symmetric subsector of the numerators. 

As a consequence of MHV helicity assignment $(1^-,2^-,3^+,\ldots,N^+)$ and the form of polarizations in eq.~(\ref{treenum5}), the only non-vanishing dot-products are $(\epsilon_2^{(-)} \cdot \epsilon_j^{(+)})$ with $j=3,4,\ldots,N$. Hence, 
all contact terms having at least two factors of $(\epsilon_i \cdot \epsilon_j)$ are bound to vanish, and crossing symmetry of the BCJ numerators is restored in the MHV sector. This is the reason why a crossing-symmetric expression (\ref{treenumA}) for the numerators w.r.t.~$2,3,\ldots,N$ emerges from their ancestors in pure-spinor superspace. On these grounds, we expect the emergence of kinematic structure constants as in eq.~(\ref{conjtree}) to extend to any multiplicity, beyond the explicit checks performed at $N\leq 6$ points.

\section{The one-loop six-point SYM amplitude}
\label{app:examples} 

The one-loop integrand (see eq.~(\ref{subamp})) for the six-point MHV single-trace subamplitude is given by
\begin{align}
&\mathcal{I}_{1,2,3,4,5,6}(\ell)=\frac{ 1 }{\ell^2 (\ell+k_1)^2} \, \Big\{ \frac{ \num_{\underline{1}|2|3|4|5|6}  }{ (\ell+k_{12})^2 (\ell+k_{123})^2 (\ell+k_{1234})^2 (\ell+k_{12345})^2} \notag \\
&\ \ \ + \frac{  \num_{\underline{1} | [2,3] |4|5|6}  }{s_{23}     (\ell+k_{123})^2 (\ell+k_{1234})^2 (\ell+k_{12345})^2} 
 + \frac{\num_{\underline{1} | 2 | [3,4] |5|6}   }{s_{34}     (\ell+k_{12})^2 (\ell+k_{1234})^2 (\ell+k_{12345})^2} \notag \\
 &\ \ \ 
  + \frac{   \num_{\underline{1} | 2 | 3| [4,5] |6}  }{s_{45}     (\ell+k_{12})^2 (\ell+k_{123})^2 (\ell+k_{12345})^2} 
   + \frac{ \num_{\underline{1} | 2 | 3 |4|  [5,6]}   }{s_{56}    (\ell+k_{12})^2 (\ell+k_{123})^2 (\ell+k_{1234})^2}  \notag \\
&\ \ \ + \Big( \frac{ X_{2,3} X_{2+3,4} }{s_{23}} +   \frac{ X_{4,3} X_{4+3,2}  }{s_{34}} \Big)  \frac{ X_{2+3+4,5} X_{2+3+4,6}  }{s_{234}   (\ell+k_{1234})^2 (\ell+k_{12345})^2}    +  \frac{ X_{2,3} X_{4,5}  \, X_{2+3,4+5} X_{2+3,6} }{s_{23} s_{45}   (\ell+k_{123})^2 (\ell+k_{12345})^2}   \notag \\
&\ \ \ + \Big( \frac{ X_{3,4} X_{3+4,5} }{s_{34}} +   \frac{ X_{5,4} X_{5+4,3}  }{s_{45}} \Big)  \frac{ X_{2,3+4+5} X_{2,6}  }{s_{345}   (\ell+k_{12})^2 (\ell+k_{12345})^2} +  \frac{ X_{2,3} X_{5,6} \, X_{2+3,4} X_{2+3,5+6} }{s_{23} s_{56}   (\ell+k_{123})^2 (\ell+k_{1234})^2} \notag \\
&\ \ \ + \Big( \frac{ X_{4,5} X_{4+5,6} }{s_{45}} +   \frac{ X_{6,5} X_{6+5,4}  }{s_{56}} \Big)  \frac{ X_{2,3} X_{2,4+5+6}  }{s_{456}   (\ell+k_{12})^2 (\ell+k_{123})^2}   +  \frac{ X_{3,4} X_{5,6}  \, X_{2,3+4} X_{2,5+6} }{s_{34} s_{56}   (\ell+k_{12})^2 (\ell+k_{1234})^2}
   \Big\} \ ,
\label{6ptex}
\end{align}
where the hexagon numerator $\num_{\underline{1}|2|3|4|5|6}$ is spelt out in eq.~(\ref{hexnum}), and the pentagon numerators are
\begin{align}
\num_{\underline{1} | [2,3] |4|5|6}  &= X_{2,3}  ( X_{2+3,5}X_{2+3,4}X_{\ell,6}+X_{2+3,6}X_{2+3,4}X_{2+3+4,5}+X_{4,6}X_{\ell,2+3}X_{2+3+4,5} )
\notag\\
\num_{\underline{1} | 2 | [3,4] |5|6}  &=
X_{3,4}  ( X_{2,5}X_{2,3+4}X_{\ell,6}+X_{2,6}X_{2,3+4}X_{2+3+4,5}+X_{3+4,6}X_{\ell,2}X_{2+3+4,5} )
\notag \\
 \num_{\underline{1} | 2 | 3| [4,5] |6} &= 
 X_{4,5} ( X_{2,4+5}X_{2,3}X_{\ell,6}+X_{2,6}X_{2,3}X_{2+3,4+5}+X_{3,6}X_{\ell,2}X_{2+3,4+5} )
\notag \\
\num_{\underline{1} | 2 | 3 |4|  [5,6]}  &=
X_{5,6} ( X_{2,4}X_{2,3}X_{\ell,5+6}+X_{2,5+6}X_{2,3}X_{2+3,4}+X_{3,5+6}X_{\ell,2}X_{2+3,4} ) \ . \label{pents}
\end{align}

\section{Checks on quadruple cuts}
\label{sec:cuts}

Here we briefly discuss some checks on quadruple cuts of SYM and supergravity amplitudes found in this paper. In addition to their string-theory derivation and BCJ duality, these checks provide more evidence for the validity of our results. To compute quadruple cuts of one-loop amplitudes~\cite{Bern:1994zx,Bern:1994cg,Britto:2004nc}, one puts four propagators on shell, i.e.~solving four equations with $K_1,K_2, K_3, K_4$ being sums of external momenta (note $K_1+K_2+K_3+K_4=0$):
\beq
\ell^2=0\,,\quad (\ell+K_1)^2=0\,,\quad (\ell+K_1+K_2)^2=0\,,\quad (\ell+K_1+K_2+K_3)^2=0 \ .
\eeq
In general there are two solutions to the set of equations, which are denoted as $\ell^{(1)}$ and $\ell^{(2)}$, and their explicit form can be found in ref.~\cite{Britto:2004nc}. Any cut of one-loop amplitudes evaluated on either of the solutions, should be equal to products of tree amplitudes summed over supermultiplets for internal lines. We work on the MHV case where the cuts are particularly simple: the only non-trivial cuts are two-mass-easy ones (including degenerate one-mass cases and, for four points, zero-mass), with either $K_1=k_1$, $K_3=k_a$, or $K_2=k_b$, $K_4=k_c$ for some external particles $a$ or $b,c$. Here the two solutions are parity conjugate to each other, and the cut is only non-vanishing on one of them, e.g.~$\ell^{(1)}$ (for the parity-conjugate $\overline{\rm MHV}$ amplitude, it will be non-vanishing on $\ell^{(2)}$). 

In practice we multiply the integrands by the four inverse propagators, and then evaluate the result on either of the solutions. For definiteness let us focus on the first case ($K_1=k_1,\ K_3=k_a$). For SYM amplitudes with canonical ordering, $K_2=k_2{+}\cdots{+}k_{a{-}1}$, $K_4=k_{a{+}1}{+}\cdots k_N$, the cut for $\ell^{(1)}$ must equal two times the corresponding box coefficient, which is
\beq
\left( \ell^2 (\ell{+}k_1)^2 (\ell{+}k_1{+}K_2)^2 (\ell{-}K_4)^2 I^{1-\te{loop, MHV}}_{1,2,\ldots,N} \right) |_{\ell=\ell^{(1)}}=\frac{\delta^8(Q)~[ K_2^2 K_4^2-(k_1+K_2)^2(k_1+K_4)^2 ]}{\langle 1\,2\rangle\langle 2\,3\rangle\cdots \langle N\,1\rangle} \ ,
\eeq
and for $\ell^{(2)}$ it must vanish. The cuts for supergravity amplitudes must also vanish for $\ell^{(2)}$, and evaluate to two times the box coefficient for $\ell^{(1)}$. The expressions of the box coefficients can be found in ref.~\cite{Bern:1998sv}. The results for the second case ($K_2=k_b, \ K_4=k_c$) follow by simple relabelings. 

We have checked numerically that our SYM results with $N=4,5,6$, \eqref{4ptex}, \eqref{5ptex} and \eqref{6ptex}, reproduce all correct quadruple cuts, and the same is true for supergravity results with $N=4,5$ as given by eq.~\eqref{sugra}. In addition, we have checked a subset of quadruple cuts for seven-point SYM and six-point supergravity 
amplitudes. 

A particularly simple case, where the computation can be done analytically, is the one-mass cut with massless corners $K_1=k_1$ and e.g.~$K_2=k_2$, $K_3=k_3$. The two solutions are
\beq
\ell^{(1)}=-| 1 \rangle \left([ 1 |+\frac{\langle 2\,3\rangle}{\langle 1\,3\rangle} [ 2 |\right),\quad \ell^{(2)}=-\left( | 1 \rangle+\frac{[2\,3]}{[1\,3]} | 2 \rangle \right) [ 1 | \ ,
\eeq
and for SYM the cuts for the two solutions are given by the tree amplitude times $s_{12} s_{23} $ and $0$, respectively. Note that by eq.~\eqref{gen_result}, all $\ell$-dependence of our numerators drops out for the first solution, thus only the scalar part contributes. For the second solution everything contributes but eq.~\eqref{gen_result} also simplifies a lot. 
This one-mass cut of SYM has been checked analytically up to six points, and with some efforts one should be able to show it to all multiplicities. It would also be interesting to prove the validity of our supergravity results on this cut. 

\section{Jacobi identities for the MHV numerators}
\label{proofBCJ}

\begin{figure}[t]
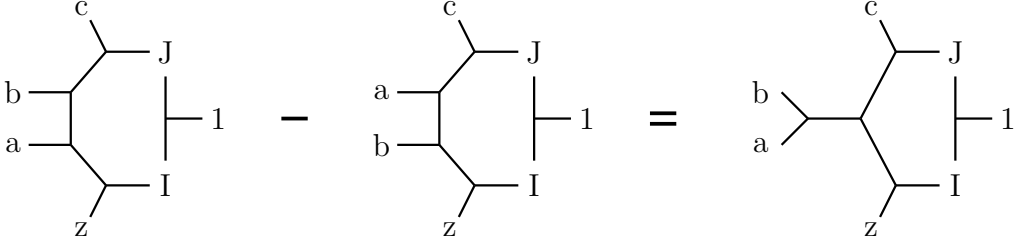

\begin{center}
 \tikzpicture [scale=0.7]
 \draw [line width=0.30mm] (-6,0) -- (-6,1);
 \draw [line width=0.30mm] (-6,1) -- (-5.33,1.77);
 \draw [line width=0.30mm] (-5.33,1.77) -- (-4.5,1.77);
 \draw [line width=0.30mm] (-6,0) -- (-5.33,-0.77);
 \draw [line width=0.30mm] (-5.33,-0.77) -- (-4.5,-0.77);
 \draw [line width=0.30mm] (-6,0) -- (-6.8,0);
 \draw [line width=0.30mm] (-6,1) -- (-6.8,1);
 \draw [line width=0.30mm] (-5.33,1.77) -- (-5.63,2.37);
 \draw [line width=0.30mm] (-5.33,-0.77) -- (-5.63,-1.37);
 \draw (-7.1,0) node {a};
 \draw (-7.1,1) node {b};
 \draw (-5.8,2.6) node {c};
 \draw (-5.8,-1.6) node {z};
 \draw (-4.2,1.77) node {J};
 \draw (-4.2,-0.77) node {I};
 \draw [line width=0.30mm] (-4.2,-0.3) -- (-4.2,1.3);
 \draw [line width=0.30mm] (-4.2,0.5) -- (-3.5,0.5);
 \draw (-3.2,0.5) node {1};
 \draw [line width=0.50mm] (-2,0.5) -- (-1.5,0.5);
 \scope[xshift=7cm]
 \draw [line width=0.30mm] (-6,0) -- (-6,1);
 \draw [line width=0.30mm] (-6,1) -- (-5.33,1.77);
 \draw [line width=0.30mm] (-5.33,1.77) -- (-4.5,1.77);
 \draw [line width=0.30mm] (-6,0) -- (-5.33,-0.77);
 \draw [line width=0.30mm] (-5.33,-0.77) -- (-4.5,-0.77);
 \draw [line width=0.30mm] (-6,0) -- (-6.8,0);
 \draw [line width=0.30mm] (-6,1) -- (-6.8,1);
 \draw [line width=0.30mm] (-5.33,1.77) -- (-5.63,2.37);
 \draw [line width=0.30mm] (-5.33,-0.77) -- (-5.63,-1.37);
 \draw (-7.1,0) node {b};
 \draw (-7.1,1) node {a};
 \draw (-5.8,2.6) node {c};
 \draw (-5.8,-1.6) node {z};
 \draw (-4.2,1.77) node {J};
 \draw (-4.2,-0.77) node {I};
 \draw [line width=0.30mm] (-4.2,-0.3) -- (-4.2,1.3);
 \draw [line width=0.30mm] (-4.2,0.5) -- (-3.5,0.5);
 \draw (-3.2,0.5) node {1};
 \draw [line width=0.50mm] (-2,0.6) -- (-1.5,0.6);
 \draw [line width=0.50mm] (-2,0.4) -- (-1.5,0.4);
\endscope
 \scope[xshift=15cm]
 \draw [line width=0.30mm] (-6,0.5) -- (-7,.5);
 \draw [line width=0.30mm] (-6,.5) -- (-5.33,1.77);
 \draw [line width=0.30mm] (-5.33,1.77) -- (-4.5,1.77);
 \draw [line width=0.30mm] (-6,.5) -- (-5.33,-0.77);
 \draw [line width=0.30mm] (-5.33,-0.77) -- (-4.5,-0.77);
 \draw [line width=0.30mm] (-7,0.5) -- (-7.5,1);
 \draw [line width=0.30mm] (-7,0.5) -- (-7.5,0);
 \draw [line width=0.30mm] (-5.33,1.77) -- (-5.63,2.37);
 \draw [line width=0.30mm] (-5.33,-0.77) -- (-5.63,-1.37);
 \draw (-7.9,0) node {a};
 \draw (-7.9,1) node {b};
 \draw (-5.8,2.6) node {c};
 \draw (-5.8,-1.6) node {z};
 \draw (-4.2,1.77) node {J};
 \draw (-4.2,-0.77) node {I};
 \draw [line width=0.30mm] (-4.2,-0.3) -- (-4.2,1.3);
 \draw [line width=0.30mm] (-4.2,0.5) -- (-3.5,0.5);
 \draw (-3.2,0.5) node {1};
\endscope
 \endtikzpicture
\end{center}
\caption{Jacobi identity involving a loop propagator.}
\label{figapp}
\end{figure}

In this appendix, we describe in detail the Jacobi identities satisfied by our one-loop MHV numerators and give a proof of eq.~(\ref{easyjac}). A Jacobi identity involves four lines (internal or external) connected by a propagator, and the three diagrams in the identity correspond to the three different channels that the propagator may represent (see figure \ref{fig4}). Now, a generic one-loop diagram is a $p$-gon with a tree attached to each corner. As elaborated in section \ref{jactypes}, it is helpful to distinguish Jacobi identities which affect propagators in tree-level subdiagrams from those relating $p$-gons and $(p\pm1)$-gons as in figure \ref{figapp}. For the former type, the Jacobi identities follow trivially from the fact that $X$ is a spinor bracket, as in the self-dual theory, since our prescription $\mathscr{X}$ preserves the $X$-structure of the trees. For the latter type, however, the Jacobi identities are non-trivial because the prescription $\mathscr{X}$ changes the $X$-structure of the $p$-gon, and we focus on this type of identity. 

As a warm up, let us first consider the one-loop all-plus numerators, which have the form
\begin{equation}
n^{1-\text{loop}}_{\text{all-plus}} = 2 \left(\prod_{i=1}^N \frac{1}{\langle \eta i \rangle^2} \right)
\times
\left( \prod_{a=1}^p X_{\ell + 1 +\cdots+(a-1),a} \right) 
\times \left( \prod_{b=1}^p X^{(A_b)} \right) \ ,
\end{equation}
where $1,2,\ldots,a$ denote the momenta entering the corners of the $p$-gon, rather than the external particles. There is a factor $X$ for each vertex in the $p$-gon, while the factors of $X^{(A_b)}$ defined in eq.~(\ref{moremass}) capture the trees. Defining the $p$-gon contribution as
\begin{equation}
x^{\text{all-plus}} = \prod_{a=1}^p X_{\ell + 1 +\cdots+(a-1),a} \ ,
\end{equation}
the Jacobi identity in figure \ref{figapp} follows from
\begin{align}
x^{\text{all-plus}}_{\cdots|a|b|\cdots} - x^{\text{all-plus}}_{\cdots|b|a|\cdots} & = 
\cdots ( X_{\ell+I+z,a} \, X_{\ell+I+z+a,b} - X_{\ell+I+z,b} \, X_{\ell+I+z+b,a} ) \cdots = \nonumber \\
& = \cdots X_{a,b}\, X_{\ell+I+z,a+b} \cdots = x^{\text{all-plus}}_{\cdots|[a,b]|\cdots}  \ .
\end{align}

In the MHV case, we consider the much more elaborate object
\begin{align}
x^{\text{MHV}} & = \mathscr{X}_{\underline{1}}\left\{ \prod_{a=1}^p X_{\ell + 1 +\cdots+(a-1),a} \right\} = \nonumber \\
& = \sum_{r<s-1}
X_{r,s}
\left(\prod_{i=2}^{r-1} X_{\ell+1+\cdots+(i-1),i}\right)
\left(\prod_{j=r+1}^{s-1} X_{1+\cdots+(j-1),j}\right)
\left(\prod_{k=s+1}^{N} X_{\ell+1+\cdots+(k-1),k}\right) \ ,
\end{align}
where $X$ is now defined with $|\eta\rangle=|1\rangle$ .
The $X_{r,s}$ factor is `non-local' in the $p$-gon, involving non-neighbouring corners, so we have to analyze how the terms with different $(r,s)$ play out together in a given Jacobi identity. We will use the notation of figure \ref{figapp}. Particle 1 must be directly attached to the $p$-gon, and plays the role of its boundary. The set $I$ contains the corners from particle 1 to $z-1$ ($I_z$ contains also $z$), and the set $J$ contains the corners from $c+1$ to particle 1 ($J_c$ contains also $c$). The terms in the Jacobi identity of figure \ref{figapp} then split according to how $(r,s)$ relate to $(a,b)$. Showing how the identity follows, the simplest cases are:

\vspace{.2cm}
$\bullet \quad r,s \in I_z \quad\textrm{or} \quad r,s \in J_c$
\vspace{-.2cm}
\beq
\cdots ( X_{\ell+I+z,a} \, X_{\ell+I+z+a,b} - X_{\ell+I+z,b} \, X_{\ell+I+z+b,a} ) \cdots =\cdots X_{a,b}\, X_{\ell+I+z,a+b} \cdots \quad \surd
\eeq

$\bullet\quad r \in I_z, \; s \in J_c$
\vspace{-.2cm}
\beq
\cdots ( X_{I+z,a} \, X_{I+z+a,b} - X_{I+z,b} \, X_{I+z+b,a} ) \cdots =\cdots X_{a,b}\, X_{I+z,a+b} \cdots
 \quad \surd
\eeq

\noindent When $r$ and $s$ coincide with either $a$ or $b$, it gets more complicated, and several terms must play out together. We now have to distinguish $(r_1,s_1)$ and $(r_2,s_2)$, where the label refers to the first and second diagrams of figure \ref{figapp}.

\vspace{.2cm}
$\bullet \quad (r_1,s_1)=(a,c),\;(r_2,s_2)=(b,c)$
\vspace{-.2cm}
\beq
\cdots ( X_{a,c} \, X_{I+z+a,b} - X_{b,c} \, X_{I+z+b,a} ) \cdots =\cdots X_{a,b}\, X_{I+z+a+b,c} \cdots
\eeq

$\bullet \quad (r_1,s_1)=(z,b),\;(r_2,s_2)=(z,a)$
\vspace{-.2cm}
\beq
\cdots ( X_{z,b} \, X_{I+z,a} - X_{z,a} \, X_{I+z,b} ) \cdots =\cdots X_{a,b}\, X_{I,z} \cdots
\eeq

$\bullet \quad r_1\in I,\;s_1=a,\quad r_2\in I,\;s_2=b$
\vspace{-.2cm}
\begin{align}
\sum_{r\in I} \cdots (X_{r,a}\,X_{\ell+I+z+a,b}-X_{r,b}\,X_{\ell+I+z+b,a}) \cdots = 
\sum_{r\in I} \cdots X_{a,b} \,X_{r,\ell+I+z+a+b} \cdots
\end{align}

$\bullet \quad r_1\in I,\;s_1=a,\quad r_2\in I,\;s_2=b$
\vspace{-.2cm}
\begin{align}
\sum_{r\in I} \cdots (X_{r,a}\,X_{\ell+I+z+a,b}-X_{r,b}\,X_{\ell+I+z+b,a}) \cdots = 
\sum_{r\in I} \cdots X_{a,b} \,X_{r,\ell+I+z+a+b} \cdots
\end{align}

$\bullet \quad r_1\in I,\;s_1=b,\quad r_2\in I,\;s_2=a$
\vspace{-.2cm}
\begin{align}
\sum_{r\in I} \cdots (X_{r,b}\,X_{I+z,a}-X_{r,a}\,X_{I+z,b}) \cdots = 
\sum_{r\in I} \cdots X_{a,b} \,X_{I+z,r} \cdots
\end{align}

$\bullet \quad r_1=b,\;s_1\in J,\quad r_2=a,\;s_2\in J$
\vspace{-.2cm}
\begin{align}
\sum_{s\in J} \cdots (X_{b,s}\,X_{\ell+I+z,a}-X_{a,s}\,X_{\ell+I+z,b}) \cdots = 
\sum_{s\in J} \cdots X_{a,b} \,X_{s,\ell+I+z} \cdots
\end{align}

$\bullet \quad r_1=a,\;s_1\in J,\quad r_2=b,\;s_2\in J$
\vspace{-.2cm}
\begin{align}
\sum_{s\in J} \cdots (X_{a,s}\,X_{I+z+a,b}-X_{b,s}\,X_{I+z+b,a}) \cdots = 
\sum_{s\in J} \cdots X_{a,b} \,X_{I+z+a+b,s} \cdots
\end{align}

\noindent The trees $a$ and $b$ form a single tree in the third diagram of figure \ref{figapp}, and a small calculation shows that the Jacobi identity requires
\begin{align}
\sum^p_{c=2}X_{\ell,c}
\left(\prod_{d=2}^{c-1} X_{\ell+2+\cdots+(d-1),d}\right)
\left(\prod_{e=c+1}^{p} X_{2+3+\cdots+(e-1),e}\right)
= \prod_{c=2}^{p} X_{\ell+2+\cdots+(c-1),c} \ .
\end{align}
This identity can be easily proven by induction.

We have completed the proof of a generic Jacobic identity. There is a more elaborate special case, where particle 1 is one of the corners involved in the relation (the third diagram vanishes then, as particle 1 would be directly attached to the loop). We have made numerical checks for the key identity (\ref{hardjac}) governing this case and will not attempt to prove it here.

\linespread{1.1}

\end{document}